\pdfoutput=1

\documentclass[11pt,twoside,a4paper,cmspaper,final,collab]{cms-tdr}
\def\svnVersion{481402P}\def\cmsCernNoTag{CERN-EP-2018-093}\def\cmsCernDate{\today}\def\cmsMessage{Published in the Journal of High Energy Physics as \href{http://dx.doi.org/10.1007/JHEP11(2018)042}{\doi{10.1007/JHEP11(2018)042.}}}

\begin{document}\cmsNoteHeader{EXO-17-023}

\hyphenation{had-ron-i-za-tion}
\hyphenation{cal-or-i-me-ter}
\hyphenation{de-vices}
\RCS$HeadURL: svn+ssh://svn.cern.ch/reps/tdr2/papers/EXO-17-023/trunk/EXO-17-023.tex $
\RCS$Id: EXO-17-023.tex 475098 2018-09-15 17:28:36Z kakwok $

\providecommand{\CL}{CL\xspace}
\newcommand{\MBH}{\ensuremath{{M_\mathrm{BH}}}\xspace}
\newcommand{\MSB}{\ensuremath{{M_\mathrm{SB}}}\xspace}
\newcommand{\MBHmin}{\ensuremath{{M_\mathrm{BH}^\text{min}}}\xspace}
\newcommand{\STmin}{\ensuremath{{S_\mathrm{T}^\text{min}}}\xspace}
\newcommand{\ST}{\ensuremath{{S_\mathrm{T}}}\xspace}
\newcommand{\MPl}{\ensuremath{{M_\mathrm{Pl}}}\xspace}
\newcommand{\NCS}{\ensuremath{{N_\mathrm{CS}}}\xspace}
\newcommand{\ES}{\ensuremath{{E_\mathrm{sph}}}\xspace}
\newcommand{\ilum}{\ensuremath{35.9\fbinv}\xspace}
\newcommand{\MS}{\ensuremath{{M_\mathrm{S}}}\xspace}
\newcommand{\RS}{\ensuremath{{R_\mathrm{S}}}\xspace}
\newcommand{\gs}{\ensuremath{{g_\mathrm{S}}}\xspace}
\newcommand{\BLACKMAX}{{\textsc{BlackMax}}\xspace}
\newcommand{\Nmin}{\ensuremath{{N^\mathrm{min}}}\xspace}
\newcommand{\nED}{\ensuremath{{n_\mathrm{ED}}}\xspace}
\newcommand{\MSgs}{\ensuremath{{\MS/\gs}}\xspace}
\newcommand{\MSgss}{\ensuremath{{\MS/\gs^2}}\xspace}
\renewcommand*{\thefootnote}{\fnsymbol{footnote}}
\providecommand{\NA}{\ensuremath{\text{---}}}
\providecommand{\cmsTable}[1]{\resizebox{\textwidth}{!}{#1}}

\cmsNoteHeader{AN-16-467}
\title{Search for black holes and sphalerons in high-multiplicity final states in proton-proton collisions at $\sqrt{s} = 13\TeV$\footnote[2]{We dedicate this paper to the memory of Prof. Stephen William Hawking, on whose transformative ideas much of this work relies.}}

\author{CMS Collaboration}
\date{\today}

\abstract{
A search in energetic, high-multiplicity final states for evidence of physics beyond the standard model, such as black holes, string balls, and electroweak sphalerons, is presented. The data sample corresponds to an integrated luminosity
of \ilum collected with the CMS experiment at the LHC in proton-proton collisions at a center-of-mass energy of 13\TeV in 2016. Standard model backgrounds, dominated by multijet production, are determined from control
regions in data without any reliance on simulation. No evidence for excesses above the predicted background is observed. Model-independent 95\% confidence level upper limits on the cross section of beyond the standard model signals
in these final states are set and further interpreted in terms of limits on semiclassical black hole, string ball, and sphaleron production. In the context of models
with large extra dimensions, semiclassical black holes with minimum masses as high as 10.1\TeV and string balls with masses as high as 9.5\TeV are excluded by this search. Results of the first dedicated search for electroweak sphalerons are presented. An upper limit of 0.021 is set at 95\% confidence level on the fraction of all quark-quark interactions above the nominal threshold energy of 9\TeV resulting in the sphaleron transition.}

\hypersetup{%
pdfauthor={CMS Collaboration},%
pdftitle={Search for black holes and sphalerons in high-multiplicity final states in proton-proton collisions at sqrt(s) = 13 TeV},%
pdfsubject={CMS},%
pdfkeywords={CMS, sphalerons, black holes}}

\maketitle
\setcounter{footnote}{0}

\section{Introduction\label{s:intro}}

Many theoretical models of physics beyond the standard model (SM)~\cite{SM1,SM2,SM3} predict strong production of particles decaying into high-multiplicity final states, i.e., characterized by three or more energetic jets, leptons, or photons. Among these models are supersymmetry~\cite{SUSY1,SUSY2,SUSY3,SUSY4,SUSY5,SUSY6,SUSY7,SUSY8}, with or without $R$-parity violation~\cite{R-parity}, and models with low-scale quantum gravity~\cite{add1,add2,add3,RS1,RS2}, strong dynamics, or other nonperturbative physics phenomena. While the final states predicted in these models differ significantly in the type of particles produced, their multiplicity, and the transverse momentum imbalance, they share the common feature of a large number of energetic objects (jets, leptons, and/or photons) in the final state. The search described in this paper targets these models of beyond-the-SM (BSM) physics by looking for final states of various inclusive multiplicities featuring energetic objects. Furthermore, since such final states can be used to test a large variety of models, we provide model-independent exclusions on hypothetical signal cross sections. Considering concrete examples of such models, we interpret the results of the search explicitly in models with microscopic semiclassical black holes (BHs) and string balls (SBs), as well as in models with electroweak (EW) sphalerons. These examples are discussed in detail in the rest of this section.

\subsection{Microscopic black holes\label{s:introBH}}

In our universe, gravity is the weakest of all known forces. Indeed, the Newton constant, ${\sim}10^{-38}\GeV^{-2}$, which governs the strength of gravity, is much smaller than the Fermi constant,  ${\sim}10^{-4}\GeV^{-2}$, which characterizes the strength of EW interactions. Consequently, the Planck scale $\MPl \sim 10^{19}\GeV$, \ie, the energy at which gravity is expected to become strong, is 17 orders of magnitude higher than the EW scale of ${\sim}100\GeV$. With the discovery of the Higgs boson~\cite{Higgs1,Higgs2,Higgs3} with a mass~\cite{H-mass,H-mass-new} at the EW scale, the large difference between the two scales poses what is known as the hierarchy problem~\cite{hierarchy}. This is because in the SM, the Higgs boson mass is not protected against quadratically divergent quantum corrections and---in the absence of fine tuning---is expected to be naturally at the largest energy scale of the theory: the Planck scale. A number of theoretical models have been proposed that attempt to solve the hierarchy problem, such as supersymmetry, technicolor~\cite{TC}, and, more recently, theoretical frameworks based on extra dimensions in space: the Arkani-Hamed, Dimopoulos, and Dvali (ADD) model~\cite{add1,add2,add3} and the Randall--Sundrum model~\cite{RS1,RS2}.

In this paper, we look for the manifestation of the ADD model that postulates the existence of $\nED \ge 2$ ``large" (compared to the inverse of the EW energy scale) extra spatial dimensions, compactified on a sphere or a torus, in which only gravity can propagate. This framework allows one to elude the hierarchy problem by explaining the apparent weakness of gravity in the three-dimensional space via the suppression of the fundamentally strong gravitational interaction by the large volume of the extra space. As a result, the fundamental Planck scale, \MD, in $3+{\nED}$ dimensions is related to the apparent Planck scale in 3 dimensions via Gauss's law as: $\MPl^2 \sim \MD^{\nED+2}R^\nED$, where $R$ is the radius of extra dimensions. Since \MD could be as low as a few TeV, \ie, relatively close to the EW scale, the hierarchy problem would be alleviated.

At high-energy colliders, one of the possible manifestations of the ADD model is the formation of microscopic BHs~\cite{dl,gt} with a production cross section proportional to the squared Schwarzschild radius, given as:
\begin{linenomath}
\begin{equation*}
\RS = \frac{1}{\sqrt{\pi}\MD}\left[\frac{\MBH}{\MD}\left(\frac{8\Gamma(\frac{\nED+3}{2})}{\nED+2}\right)\right]^{\frac{1}{\nED+1}},
\end{equation*}
\end{linenomath}
where $\Gamma$ is the gamma function and \MBH is the mass of the BH. In the simplest production scenario, the cross section is given by the area of a disk of radius \RS,
\ie, $\sigma \approx \pi \RS^2$~\cite{dl,gt}. In more complicated production scenarios, \eg, a scenario with energy loss during the formation of the BH horizon, the cross section is modified from this ``black disk" approximation by a factor of order one~\cite{gt}.

As BH production is a threshold phenomenon, we search for BHs above a certain minimum
mass $\MBHmin \ge \MD$. In the absence of signal, we will express the results of the search as limits on \MBHmin.
In the semiclassical case (strictly valid for $\MBH \gg \MD$), the BH quickly evaporates via Hawking radiation~\cite{Hawking} into
a large number of energetic particles, such as gluons, quarks, leptons, photons, \etc The relative abundance of various particles produced in the process of BH evaporation is expected
to follow the number of degrees of freedom per particle in the SM. Thus, about 75\% of particles produced are expected to be
quarks and gluons, because they come in three or eight color combinations, respectively. A significant amount of missing
transverse momentum may be also produced in the process of BH evaporation via production
of neutrinos, which constitute ${\sim}5\%$ of the products of a semiclassical BH decay, \PW\ and \PZ\ boson decays, heavy-flavor quark decays, gravitons, or noninteracting stable BH remnants.

If the mass of a BH is close to \MD, it is expected to exhibit quantum features, which can modify the characteristics of its decay products. For example, quantum BHs~\cite{QBH1,QBH2,QBH3} are expected to decay before they thermalize, resulting in low-multiplicity final states. Another model of semiclassical BH precursors is the SB model~\cite{sb}, which predicts the formation of a long, jagged string excitation, folded into a ``ball". The evaporation of an SB is similar to that of a semiclassical BH, except that it takes place at a fixed Hagedorn temperature~\cite{Hagedorn}, which depends only on the string scale \MS. The formation of an SB occurs once the mass of the object exceeds \MSgs, where \gs is the string coupling. As the mass of the SB grows, eventually it will transform into a semiclassical BH, once its mass exceeds $\MSgss > \MD$.

A number of searches for both semiclassical and quantum BHs, as well as for SBs have been performed at the CERN LHC using the Run 1 ($\sqrt{s} = 7$ and 8\TeV) and Run 2 ($\sqrt{s} = 13\TeV$) data. An extensive review of Run 1 searches can be found in Ref.~\cite{GL-Springer}. The most recent Run 2 searches for semiclassical BHs and SBs were carried out by ATLAS~\cite{ATLAS-inclusive13,Aaboud:2016ewt} and CMS~\cite{CMSBH4} using 2015 data. Results of searches for quantum BHs in Run 2 based on 2015 and 2016 data can be found in Refs.~\cite{ATLAS:2015nsi,Aad:2015ywd,Sirunyan:2017ygf,Aaboud:2017yvp,Aaboud:2017nak,Sirunyan:2018zhy}. The most stringent limits on \MBHmin set by the Run 2 searches are 9.5 and 9.0 TeV for semiclassical and quantum BHs, respectively,  for $\MD = 4\TeV$~\cite{ATLAS-inclusive13,CMSBH4}. The analogous limits on the minimum SB mass depend on the choice of the string scale and coupling and are in the 6.6--9\TeV range for the parameter choices considered in Refs.~\cite{ATLAS-inclusive13,CMSBH4}.

\subsection{Sphalerons\label{s:introSph}}

The Lagrangian of the EW sector of the SM has a possible nonperturbative solution,
which includes a vacuum transition known as a ``sphaleron". This class of solutions to gauge field theories was first proposed in 1976 by 't~Hooft \cite{tHooft:76}. The particular sphaleron solution of the SM was first described by Klinkhamer and Manton in 1984 \cite{Manton:84}. It is also
a critical piece of EW baryogenesis theory~\cite{Trodden:99}, which explains
the matter-antimatter asymmetry of the universe by such processes. The crucial feature of the sphaleron, which allows such claims to
be made, is the violation of baryon ($B$) and lepton ($L$) numbers, while preserving $B-L$. The possibility of sphaleron transitions at hadron colliders and related phenomenology has been discussed since the late 1980s~\cite{Ringwald:1989ee}.

Within the framework of perturbative SM physics, there are twelve globally conserved currents, one for each of the 12 fundamental fermions: $J^\mu = \overline{\psi}_L \gamma^\mu \psi_L$. An anomaly breaks this conservation, in particular $\partial_\mu J^\mu = [g^2/(16 \pi^2)] \mathrm{Tr}[F_{\mu\nu} \widetilde{F}^{\mu\nu}]$. This is because the integral of this term, known as a Chern--Simons (or winding) number $\NCS$~\cite{CSnumber}, is nonzero. The anomaly exists for each fermion doublet. This means that the lepton number changes by 3\NCS, since each of three leptons produced has absolute lepton number of 1. The baryon number will also change by $3 \NCS$ because each quark has an absolute baryon number of $1/3$ and there are three colors and three generations of quarks produced. This results in two important relations, which are essential to the phenomenology of sphalerons: $\Delta(B+L) = 6 \NCS$ and $\Delta(B-L) = 0$. The anomaly only exists if there is enough energy to overcome the potential in \NCS, which is fixed by the values of the
EW couplings. Assuming the state at 125 GeV to be the SM Higgs boson, the precise measurement of its mass~\cite{H-mass,H-mass-new}
allowed the determination of these couplings, giving an estimate of the energy required for the sphaleron transitions of $\ES \approx 9\TeV$~\cite{Manton:84,TyeWong}.

While the $\ES$ threshold is within the reach of the LHC, it was originally thought that the sphaleron transition probability would be significantly suppressed by a large potential barrier. However, in a recent work~\cite{TyeWong} it has been suggested that the periodic nature of the Chern--Simons potential reduces this suppression at collision energies $\sqrt{\hat s} < \ES$, removing it completely for $\sqrt{\hat s} \ge \ES$. This argument opens up the possibility of observing an EW sphaleron transition  in proton-proton (\Pp\Pp) collisions at the LHC via processes such as: $\cPqu + \cPqu \to \Pep \Pgmp \Pgt^+$ $\PAQt\, \PAQt\, \PAQb\, \PAQc\, \PAQc\, \PAQs\, \PAQd + X$. Fundamentally, the $\NCS = +1$ ($-1$) sphaleron transitions involve 12 (anti)fermions: three (anti)leptons, one from each generation, and nine (anti)quarks, corresponding to three colors and three generations, with the total electric charge and weak isospin of zero. Nevertheless, at the LHC, we consider signatures with 14, 12, or 10 particles produced, that arise from a $\cPq + \cPq' \to \cPq + \cPq'  + \mathrm{sphaleron}$ process, where 0, 1, or 2 of the 12 fermions corresponding to the sphaleron transition may ``cancel" the $\cPq$ or $\cPq'$ inherited from the initial state~\cite{Ellis2016,sphaleron-gen}. Since between zero and three of the produced particles are neutrinos, and also between zero and three are top quarks, which further decay, the actual multiplicity of the visible final-state particles may vary between 7 and 20 or more. Some of the final-state particles may also be gluons from either initial- or final-state radiation. While the large number of allowed combinations of the 12 (anti)fermions results in over a million unique transitions~\cite{TEU_Kolb}, many of the final states resulting from these transitions would look identical in a typical collider experiment, as no distinction is made between quarks of the first two generations, leading to only a few dozen phenomenologically unique transitions, determined by the charges and types of leptons and the third-generation quarks in the final state. These transitions would lead to characteristic collider signatures, which would have many energetic jets and charged leptons, as well as large missing transverse momentum due to undetected neutrinos.

A phenomenological reinterpretation in terms of limits on the EW sphaleron production of an ATLAS search for microscopic BHs in the multijet final states at $\sqrt{s} = 13\TeV$~\cite{ATLAS-inclusive13}, comparable to an earlier CMS analysis~\cite{CMSBH4}, was recently performed in Ref.~\cite{Ellis2016}. In the present paper, we describe the first dedicated experimental search for EW sphaleron transitions.

\section{The CMS detector and the data sample\label{sec:detector}}

The central feature of the CMS apparatus is a superconducting solenoid of 6\unit{m} internal diameter, providing a magnetic field of 3.8\unit{T}. Within the solenoid volume are a silicon pixel and strip tracker, a lead tungstate crystal electromagnetic calorimeter (ECAL), and a brass and scintillator hadron calorimeter (HCAL), each composed of a barrel and two endcap sections. Forward calorimeters extend the pseudorapidity ($\eta$) coverage provided by the barrel and endcap detectors. Muons are detected in gas-ionization chambers embedded in the steel flux-return yoke outside the solenoid.

In the region $\abs{\eta} < 1.74$, the HCAL cells have widths of 0.087 in pseudorapidity and 0.087 in azimuth ($\phi$). In the $\eta-\phi$ plane, and for $\abs{\eta} < 1.48$, the HCAL cells map on to $5 \times 5$ arrays of ECAL crystals to form calorimeter towers projecting radially outwards from close to the nominal interaction point. For $\abs{\eta}  > 1.74$, the coverage of the towers increases progressively to a maximum of 0.174 in $\Delta \eta$ and $\Delta \phi$. Within each tower, the energy deposits in ECAL and HCAL cells are summed to define the calorimeter tower energies, subsequently used to provide the energies and directions of hadronic jets.

Events of interest are selected using a two-tiered trigger system~\cite{Khachatryan:2016bia}. The first level, composed of custom hardware processors, uses information from the calorimeters and muon detectors to select events at a rate of around 100\unit{kHz} within a time interval of less than 4\mus. The second level, known as the high-level trigger (HLT), consists of a farm of processors running a version of the full event reconstruction software optimized for fast processing, and reduces the event rate to around 1\unit{kHz} before data storage.

A more detailed description of the CMS detector, together with a definition of the coordinate system used and the relevant kinematic variables, can be found in Ref.~\cite{Chatrchyan:2008zzk}.

The analysis is based on a data sample recorded with the CMS detector in \Pp\Pp\ collisions at a center-of-mass energy of 13\TeV in 2016, corresponding to an integrated luminosity of \ilum. Since typical signal events are expected to contain multiple jets, we employ a trigger based on the \HT variable, defined as the scalar sum of the transverse momenta (\pt) of all jets in an event reconstructed at the HLT. We require $\HT > 800$--900 \GeV and also use a logical OR with several single-jet triggers with \pt thresholds of 450--500 \GeV. The resulting trigger selection is fully efficient for events that subsequently satisfy the offline requirements used in the analysis.

\section{Event reconstruction\label{sec:eventselection}}

The particle-flow (PF) algorithm~\cite{PFLOW} aims to reconstruct and identify each individual particle in an event with an optimized combination of information from the various elements of the CMS detector. The energy of photons is directly obtained from the ECAL measurement, corrected for zero-suppression effects. The energy of electrons is determined from a combination of the electron momentum at the primary interaction vertex as determined by the tracker, the energy of the corresponding ECAL cluster, and the energy sum of all bremsstrahlung photons spatially compatible with originating from the electron track. The energy of muons is obtained from the curvature of the corresponding track. The energy of charged hadrons is determined from a combination of their momentum measured in the tracker and the matching ECAL and HCAL energy deposits, corrected for zero-suppression effects and for the response function of the calorimeters to hadronic showers. Finally, the energy of neutral hadrons is obtained from the corresponding corrected ECAL and HCAL energies.

The reconstructed vertex with the largest value of summed physics-object $\pt^2$ is taken to be the primary $\Pp\Pp$ interaction vertex. The physics objects are the jets, clustered using the anti-\kt jet finding algorithm~\cite{Cacciari:2008gp,Cacciari:2011ma} with the tracks assigned to the vertex as inputs, and the associated missing transverse momentum, taken as the negative vector sum of the \pt of those jets. Events are required to have at least one reconstructed vertex within 24 (2)\unit{cm} of the nominal collision point in the direction parallel (perpendicular) to the beams.

For each event, hadronic jets are clustered from the PF candidates using the anti-\kt algorithm with a distance parameter of 0.4. The jet momentum is determined as the vectorial sum of all particle momenta in the jet, and is found from simulation to be within 5 to 10\% of the true momentum over the whole \pt spectrum and detector acceptance. Additional \Pp\Pp\  interactions within the same or neighboring bunch crossings (pileup) can contribute additional tracks and calorimetric energy depositions to the jet momentum. To mitigate this effect, tracks originating from pileup vertices are discarded and an offset correction is applied to correct for the remaining contributions. Jet energy corrections are derived from simulation, to bring the measured response of jets to that of particle-level jets on average. In situ measurements of the momentum balance in dijet, multijet, {\Pgg}+jet, and leptonically decaying {\PZ}+jet events are used to account for any residual differences in the jet energy scales in data and simulation~\cite{Khachatryan:2016kdb}. The jet energy resolution amounts typically to 15\% at a jet \pt of 10\GeV, 8\% at 100\GeV, and 4\% at 1\TeV. Additional selection criteria are applied to each jet to remove those potentially dominated by anomalous contributions from various subdetector components or reconstruction failures. All jets are required to have $\pt >  70\GeV$ and be within $\abs{\eta} < 5$. For the leading \pt jet in each event, the energy fraction carried by muon candidates failing the standard identification~\cite{Sirunyan:2018fpa} is required to be less than 80\%. This requirement removes events where a low-momentum muon is misreconstructed with very high momentum and misidentified as a high-energy jet. We further require the leading jet in an event to have a charged-hadron fraction of less than 0.99 if this jet is found within $\abs{\eta} <2.4$~\cite{CMS-PAS-JME-16-003}.

The missing transverse momentum, $\ptmiss$, is defined as the magnitude of the vectorial sum of transverse momenta of all PF candidates in an event. The jet energy corrections are further propagated to the \ptmiss calculation.

Details of muon reconstruction can be found in Ref.~\cite{Sirunyan:2018fpa}. The muon candidate is required to have at least one matching energy deposit in the pixel tracker and at least six deposits in the silicon strip tracker, as well as at least two track segments in the muon detector. The transverse impact parameter and the longitudinal distance of the track associated with the muon with respect to the primary vertex are required to be less than 2 and 5\unit{mm}, respectively, to reduce contamination from cosmic ray muons. The global track fit to the tracker trajectory and to the muon detector segments must have a $\chi^2$ per degree of freedom of less than 10. Muon candidates are required to have $\pt > 70\GeV$ and to be within $\abs{\eta} < 2.4$.

Details of electron and photon reconstruction can be found in Refs.~\cite{EGM-13-001} and \cite{EGM-14-001}, respectively. Electron and photon candidates are required to have $\pt> 70\GeV$ and $\abs{\eta} < 2.5$, excluding the $1.44 < \abs{\eta} < 1.57$ transition region between the ECAL barrel and endcap detectors where the reconstruction is suboptimal. We use standard identification criteria, corresponding to an average efficiency of 80\% per electron or photon. The identification criteria include a requirement that the transverse size of the electromagnetic cluster be compatible with the one expected from a genuine electron or photon, and that the ratio of the HCAL to ECAL energies be less then 0.25 (0.09) for electrons and less than 0.0396 (0.0219) for photons in the barrel (endcap). In addition, photon candidates are required to pass the conversion-safe electron veto requirements~\cite{EGM-14-001}, which disambiguates them from electron candidates.

Muons, electrons, and photons are required to be isolated from other energy deposits in the tracker and the calorimeters. The isolation $\mathcal{I}$  is defined as the ratio of the \pt sum of various types of additional PF candidates in a cone of radius $\Delta R = \sqrt{\smash[b]{(\Delta\eta)^2+(\Delta\phi)^2}}$ of 0.4 (muons) or 0.3 (electrons and photons), centered on the lepton or photon candidate, to the candidate's \pt. For muons, the numerator of the ratio is corrected for the contribution of neutral particles due to pileup, using one half of the \pt carried by the charged hadrons originating from pileup vertices. For electrons and photons, an average area method~\cite{rho-method}, as estimated with {\FASTJET}~\cite{Cacciari:2011ma}, is used. The isolation requirements are the same as used in an earlier 13\TeV analysis~\cite{CMSBH4}, except that for electrons we use a tighter isolation requirement of ${\cal I} < 0.07$.

To avoid double counting, we remove jets that are found within a radius of $\Delta R =0.3$ from a muon, electron, or photon, if the latter object contributes more than 80, 70, or 50\% of the jet \pt, respectively.

\section{Analysis strategy\label{s:strategy}}

We follow closely the approach for semiclassical BH searches originally developed by CMS for Run 1 analyses~\cite{CMSBH1,CMSBH2,CMSBH3} and subsequently used in the studies of early Run 2~\cite{CMSBH4} data. This approach is based on an inclusive search for BH decays to
all possible final states, dominated by the high-multiplicity multijet ones in the semiclassical BH case. This type of
analysis is less sensitive to the details of BH evaporation and the relative abundance of various particles produced, as it
considers all types of particles in the final state. We use a single discriminating variable \ST, defined as the scalar sum of
\pt of all $N$ energetic objects in an event (which we define as jets, electrons, muons, and photons with \pt
above a given threshold), plus \ptmiss in the event, if it exceeds the same threshold:
$\ST = \ptmiss + \sum_{i=1}^{N} \pt^i$. Accounting for \ptmiss in the \ST variable makes \ST a better measure of the total transverse momentum in the event carried by all the various particles. Since it is impossible to tell how many objects lead to the \ptmiss in the event, we do not consider \ptmiss values above the threshold when determining the object multiplicity.

This definition of \ST is robust against variations in the BH evaporation model, and is also sensitive to the cases when there is large
\ptmiss due to enhanced emission of gravitons or to models in which a massive, weakly interacting remnant of a BH is formed at the terminal stage of Hawking evaporation, with a mass below \MD. It is equally applicable to sphaleron searches, given the expected energetic, high-multiplicity final states, possibly with large \ptmiss.

The \ST distributions are then considered separately for various inclusive object multiplicities (\ie, $N \ge \Nmin = 3, \dots,11$).
The background is dominated by SM QCD multijet production and is estimated exclusively from control samples in data. The observed number of events with \ST values above a
chosen threshold is compared with the background and signal+background predictions to either establish a signal or to set limits on the signal production. This approach does not rely on the Monte Carlo (MC) simulation of the backgrounds, and
it also has higher sensitivity than exclusive searches in specific final states, \eg, lepton+jets~\cite{ATLAS-ljets1,ATLAS-ljets2}.

The main challenge of the search is to describe the inclusive multijet background in a robust
way, as both BH and sphaleron signals correspond to a broad enhancement in the high tail of the \ST distribution, rather than to a narrow peak.
Since these signals are expected to involve a high multiplicity
of final-state particles, one has to reliably describe the background for large jet multiplicities,
which is quite challenging theoretically as higher-order calculations that fully describe multijet
production do not exist. Thus, one cannot rely on simulation to reproduce the \ST spectrum for large $N$ correctly.

To overcome this problem, a dedicated method of predicting the QCD multijet background directly from collision
data has been developed for the original Run 1 analysis~\cite{CMSBH1} and used in the subsequent Run 1~\cite{CMSBH2,CMSBH3} and Run 2~\cite{CMSBH4} searches. It has been found empirically, first via simulation-based studies, and then from the analysis of data at low jet
multiplicities, that the shape of the \ST distribution for the dominant QCD multijet background
does not depend on the multiplicity of the final state, above a certain turn-on threshold.
This observation reflects the way a parton shower develops via nearly collinear emission, which
conserves \ST. It allows one to predict the \ST spectrum of a multijet final state using low-multiplicity
QCD events, \eg, dijet or trijet events. This ``\ST invariance" provides a powerful method of predicting the dominant
background for BH production by taking the \ST shape from low-multiplicity events, for which the signal contamination
is expected to be negligible, and normalizing it to the observed spectrum at high multiplicities at the low
end of the \ST distribution, where signal contamination is negligible even for large multiplicities of the
final-state objects. The method has been also used for other CMS searches, \eg, a search for stealth
supersymmetry~\cite{stealth} and a search for multijet resonances~\cite{EXO-13-001}.

\section{Simulated samples\label{sec:signals}}
\subsection{Black hole and string ball signal samples}

Signal simulation is performed using the \BLACKMAX~v2.02.0~\cite{BlackMax} (semiclassical BHs) and \CHARYBDIS2~v1.003~\cite{Charybdis,Charybdis2}
(semiclassical BHs and SBs)  generators. The generator settings of each model are listed in Tables \ref{table:generator-blackMax}
and \ref{table:generator-Charybdis}.

\begin{table}[hbt]
\centering
\topcaption{Generator settings used for \BLACKMAX signal sample generation.\label{table:generator-blackMax}}
\cmsTable{
\begin{tabular}{ccccc}
Model & Choose\_a\_case   & Mass\_loss\_factor & Momentum\_loss\_factor & turn\_on\_graviton\\
\hline
B1    &   tensionless\_nonrotating &  0 & 0 & FALSE\\
B2    &   rotating\_nonsplit &  0 & 0 & FALSE\\
B3    &   rotating\_nonsplit &  0.1 & 0.1 & TRUE\\
\end{tabular}
}
\end{table}

\begin{table}[hbt]
\centering
\topcaption{Generator settings used for \CHARYBDIS2 signal sample generation.\label{table:generator-Charybdis}}
\cmsTable{
\begin{tabular}{ccccccccc}
Model & BHSPIN   & MJLOST & YRCSC & NBODYAVERAGE & NBODYPHASE & NBODYVAR & RMSTAB & RMBOIL\\
\hline
C1  & TRUE  & FALSE & FALSE & FALSE & TRUE & TRUE &FALSE &FALSE\\
C2  & FALSE & FALSE & FALSE & FALSE & TRUE & TRUE &FALSE &FALSE\\
C3  & TRUE  & FALSE & FALSE & TRUE  & FALSE& FALSE &FALSE &FALSE\\
C4  & TRUE  & TRUE  & TRUE  & FALSE & TRUE & TRUE &FALSE &FALSE\\
C5  & TRUE  & TRUE  & TRUE  & FALSE & FALSE& FALSE &TRUE &FALSE\\
C6  & TRUE  & TRUE  & TRUE  & FALSE & FALSE & FALSE &FALSE &TRUE\\
\end{tabular}
}
\end{table}

For semiclassical BH signals, we explore different aspects of BH production and decay by simulating various scenarios, including
nonrotating BHs (B1,C2), rotating BHs (B2,C1), rotating BHs with mass loss (B3), and rotating BHs with Yoshino--Rychkov bounds~\cite{YR} (C4).
Models C3, C5, and C6 explore the termination phase of the BH with different object multiplicities from the BH remnant, varying from 2-body decaying remnant (C3), stable remnant (C5, for which additionally the generator parameter NBODY was changed from its default value of 2 to 0), and "boiling" remnant (C6), where the remnant continues to evaporate until a maximum Hawking
temperature equal to \MD is reached. For each model,
the fundamental Planck scale \MD is varied within 2--9\TeV in 1\TeV steps, each with $\nED=2,\,4,\,6$. The minimum black hole mass \MBHmin is varied between $\MD+1\TeV$ and 11\TeV in 1\TeV steps.

For SB signals, two sets of benchmark points are generated with \CHARYBDIS2, such that different regimes of the SB production
can be explored. For a constant string coupling value $\gs = 0.2$ the string scale \MS is varied from 2 to 4\TeV, while at constant $\MS=3.6\TeV$, \gs is varied from 0.2 to 0.4. For all SB samples, $\nED = 6$ is used. The SB dynamics below the
first transition (\MSgs), where the SB production cross section scales with $\gs^2/\MS^4$, are probed with the constant $\gs = 0.2$ and low  \MS values as well as with the constant \MS scan. The saturation regime ($\MSgs < \MSB < \MSgss$), where the SB production cross section no longer depends on \gs, is probed
by the higher \MS points of the constant \gs benchmark. For each benchmark point, the scale \MD is chosen such that the cross section
at the SB--BH transition (\MSgss) is continuous.

For the BH and SB signal samples we use leading order (LO) MSTW2008LO~\cite{MSTW,MSTW1} parton distribution functions (PDFs).
This choice is driven by the fact that this set tends to give a conservative estimate of the signal cross section at
high masses, as checked with the modern  NNPDF3.0~\cite{NNPDF} LO PDFs, with the value of strong coupling constant
of 0.118 used for the central prediction, with a standard uncertainty eigenset. The MSTW2008LO PDF set was also used
in all Run 1 BH searches~\cite{CMSBH1,CMSBH2,CMSBH3} and in an earlier Run 2~\cite{CMSBH4} search, which makes the comparison with earlier results straightforward.

\subsection{Sphaleron signal samples\label{sec:sphaleron-signal}}

The electroweak sphaleron processes are generated at LO with the \textsc{BaryoGEN} v1.0 generator~\cite{sphaleron-gen}, capable of simulating various final states described in Section~\ref{s:introSph}. We simulate the sphaleron signal for three values of the transition energy $\ES = 8$, 9, and 10\TeV. The parton-level simulation is done with the CT10 LO PDF set~\cite{CT10}. In the process of studying various PDF sets, we found that the NNPDF3.0 yields a significantly larger fraction of sea quarks in the kinematic region of interest than all other modern PDFs. While the uncertainty in this fraction is close to 100\%, we chose the CT10 set, for which this fraction is close to the median of the various PDF sets we studied. The PDF uncertainties discussed in Section~\ref{s:systematics} cover the variation in the signal acceptance between various PDFs due to this effect.

\begin{figure}[htb]
    \centering
    \includegraphics[width=0.45\textwidth]{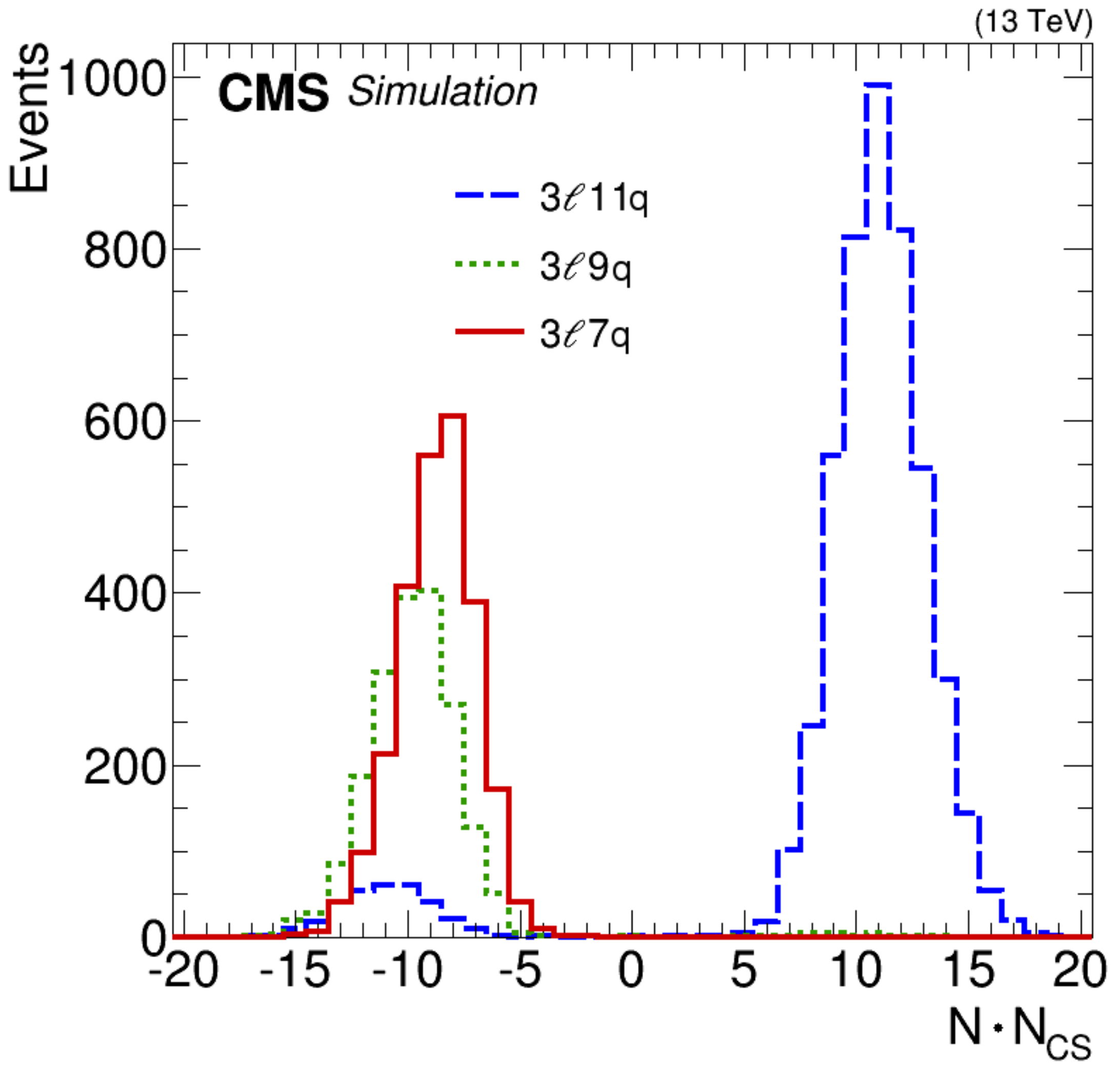}
    \caption{Observed final-state particle multiplicity $N$ distributions for $\NCS=\pm 1$ sphaleron transitions resulting in 10, 12, and 14 parton-level final-state multiplicities. The relative numbers of events in the  histograms are proportional to the relative probabilities of these three parton-level configurations. The peaks at positive values correspond to $\NCS = 1$ transitions, while those at negative values correspond to $\NCS = -1$ transitions and therefore are shifted toward lower multiplicity $N$ because of cancellations with initial-state partons.}
    \label{fig:sphaleron}
\end{figure}

The typical final-state multiplicities for the $\NCS = \pm1$ sphaleron transitions resulting in 10, 12, or 14 parton-level final states are shown in Fig.~\ref{fig:sphaleron}. The $\NCS = 1$ transitions are dominated by 14 final-state partons, as the proton mainly consists of valence quarks, thus making the probability of cancellations small.

The cross section for sphaleron production is given by~\cite{Ellis2016}: $\sigma  = \mathrm{PEF} \, \sigma_0$, where $\sigma_0 = 121,$ 10.1, and 0.51\unit{fb} for $\ES = 8,$ 9, and 10\TeV, respectively, and PEF is the pre-exponential factor, defined as the fraction of all quark-quark interactions above the sphaleron energy threshold \ES that undergo the sphaleron transition.

\subsection{Background samples}

In addition, we use simulated samples of {\PW}+jets, {\PZ}+jets, {\Pgg}+jets, \ttbar, and QCD multijet events for auxiliary studies. These events are generated with the  \MGvATNLO v2.2.2~\cite{MadGraph} event generator at LO or next-to-LO, with the NNPDF3.0 PDF set of a matching order.

The fragmentation and hadronization of parton-level signal and background  samples is done with \PYTHIA v8.205~\cite{pythia8}, using the underlying event tune CUETP8M1~\cite{CUET}.  All signal and background samples are reconstructed with the detailed simulation of the CMS detector via \GEANTfour~\cite{GEANT4}. The effect of pileup interactions is simulated by superimposing simulated minimum bias events on the hard-scattering interaction, with the multiplicity distribution chosen to match the one observed in data.

\section{Background estimate\label{s:backgrounds}}
\subsection{Background composition}

\begin{figure}[htb]
    \centering
    \includegraphics[width=0.47\textwidth]{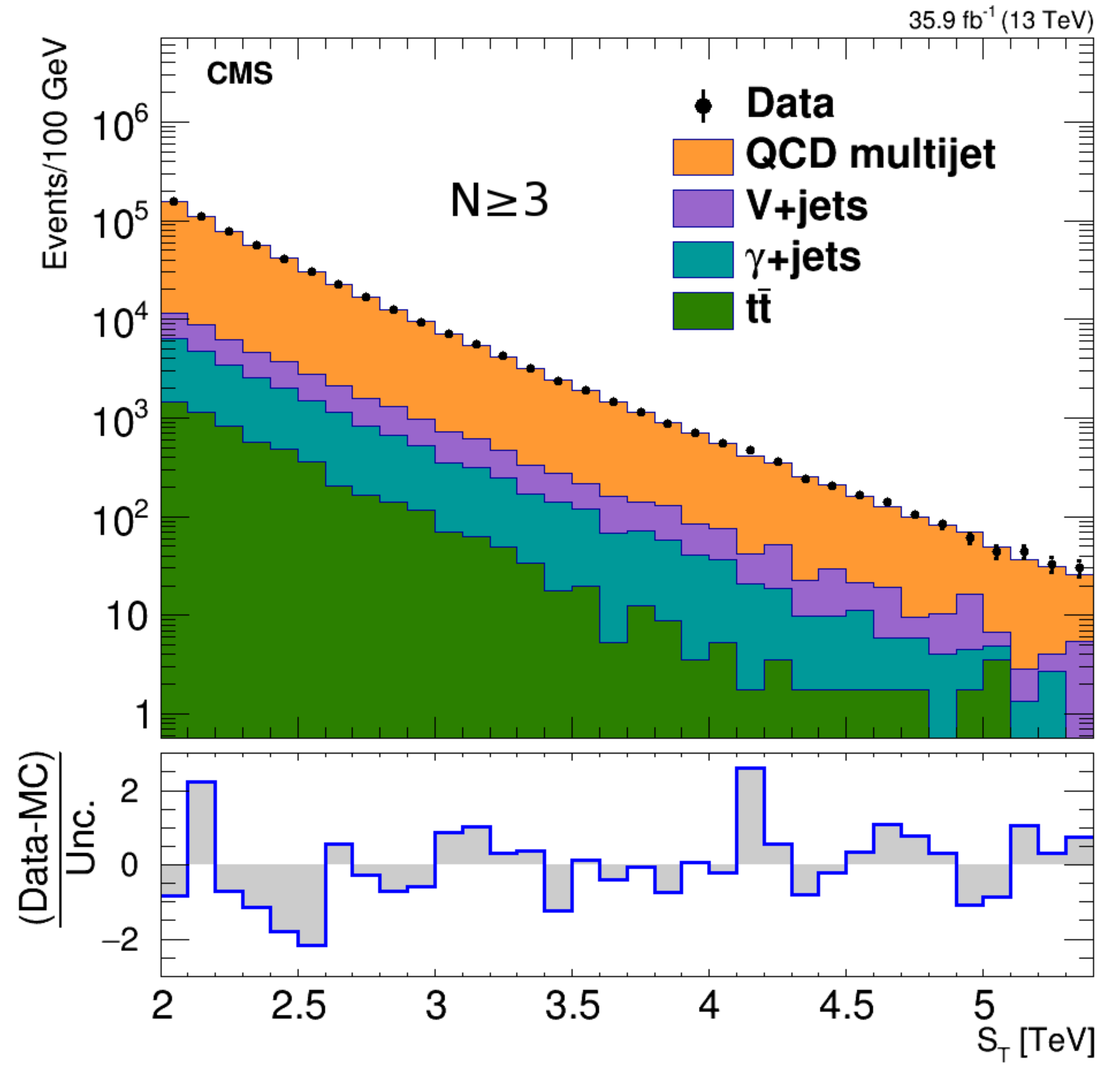}
    \includegraphics[width=0.47\textwidth]{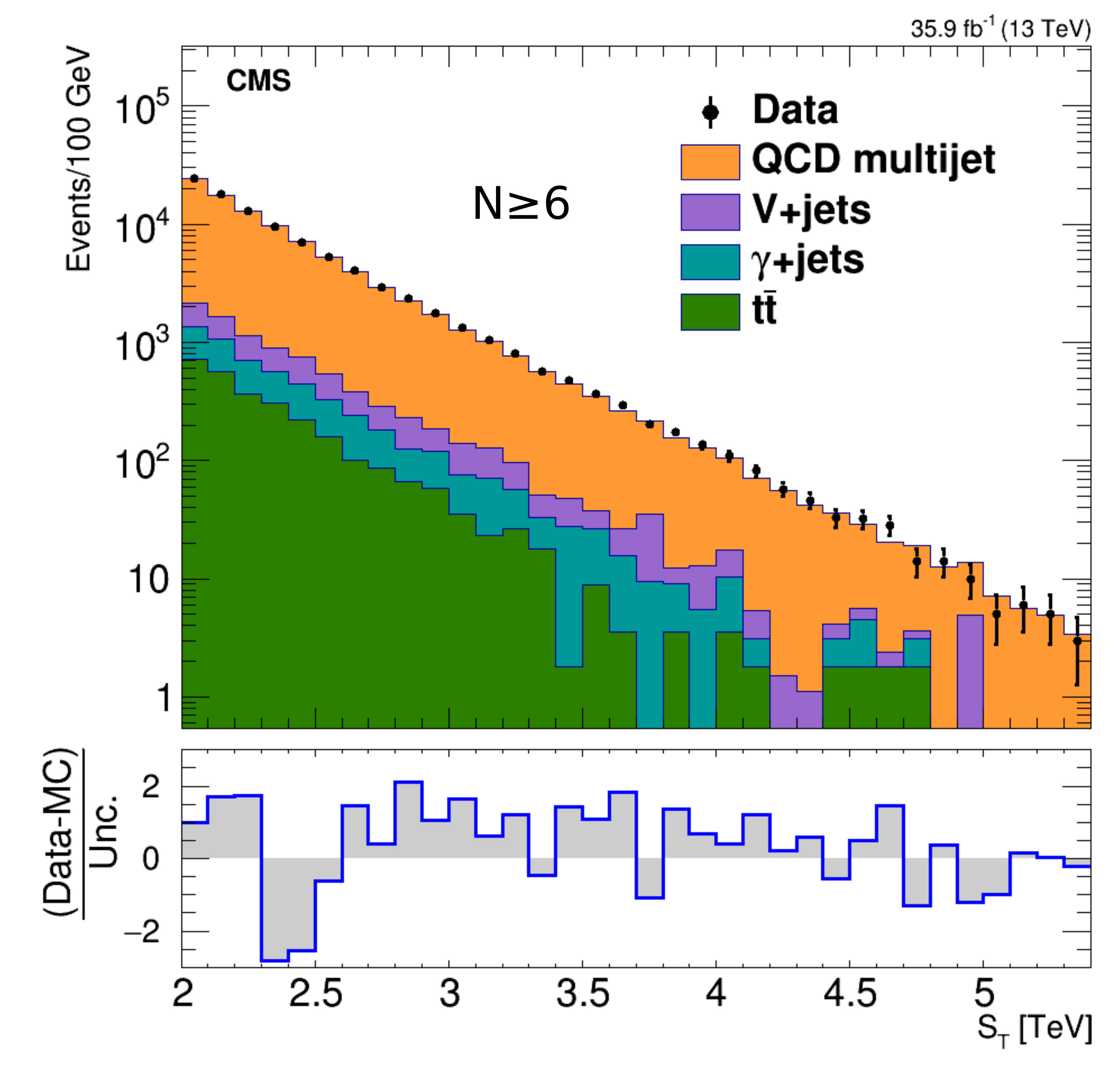}
    \caption{The \ST distribution in data for inclusive multiplicities of (left) $N \ge 3$ and (right) $N \ge 6$, compared with the normalized background prediction from simulation, illustrating the relative contributions of major backgrounds. The lower panels show the difference between the data and the simulated background prediction, divided by the statistical uncertainty in data. We note that despite an overall agreement, we do not rely on simulation for obtaining the background prediction.}
    \label{fig:STdist}
\end{figure}

The main backgrounds in the analyzed multi-object final states are: QCD multijet, V+jets (where V = \PW, \PZ), \Pgg+jets, and \ttbar production, with the QCD multijet background being by far the most dominant.
Figure~\ref{fig:STdist} illustrates the relative importance of these backgrounds for the inclusive multiplicity $N \ge 3$ and 6 cases, based on simulated background samples. To reach the overall agreement with the data, all simulated backgrounds except for the QCD multijets are normalized to the most accurate theoretical predictions available, while the QCD multijet background is normalized so that the total number of background events matches that in data. While we do not use simulated backgrounds to obtain the main results in this analysis, Fig.~\ref{fig:STdist} illustrates an important point: not only is the QCD multijet background at least an order of magnitude more important than other backgrounds, for both low- and high-multiplicity cases, but also the shape of the \ST distributions for all major backgrounds is very similar, so the method we use to estimate the multijet background, discussed below, provides an acceptable means of predicting the overall background as well.

\subsection{Background shape determination}

The background prediction method used in the analysis follows  closely that in previous similar CMS searches~\cite{CMSBH1,CMSBH2,CMSBH3,CMSBH4}.
As discussed in Section~\ref{s:strategy}, the central idea of this method is that the shape of the \ST distribution for the dominant multijet background is invariant with respect to the final-state object multiplicity $N$. Consequently, the background shape can be extracted from low-multiplicity spectra and used to describe the background at high multiplicities. The \ST value is preserved by the final-state radiation, which is the dominant source of extra jets beyond LO $2 \to 2$ QCD processes, as long as the additional jets are above the \pt threshold used in the definition of \ST.
At the same time, jets from initial-state radiation (ISR) change the \ST value, but because their \pt spectrum is steeply falling they typically contribute only a few percent to the \ST value and change the multiplicity $N$ by just one unit, for events used in the analysis. Consequently, we extract the background shape from the $N=3$ \ST spectrum, which already has a contribution from ISR jets, and therefore reproduces the \ST shape at higher multiplicities better than the $N=2$ spectrum used in earlier analyses. To estimate any residual noninvariance in the \ST distribution, the $N=4$ \ST spectrum, normalized to the $N=3$ spectrum in terms of the total number of events, is also used as an additional component of the background shape uncertainty. Furthermore, to be less sensitive to the higher instantaneous luminosity delivered by the LHC in 2016, which resulted in a higher pileup, and to further reduce the effect of ISR, the \pt threshold for all objects was raised to 70\GeV, compared to 50\GeV used in earlier analyses. The reoptimization that has resulted in the choice of a new exclusive multiplicity to be used for the baseline QCD multijet background prediction and a higher minimum \pt threshold for the objects counted toward \ST was based on extensive studies of MC samples and low-\ST events in data.

In order to obtain the background template, we use a set of 16 functions employed in earlier searches for BSM physics in dijets, VV events, and multijet events at various colliders. These functions typically have an exponential or power-law behavior with \ST, and are described by 3--5 free parameters. Some of the functions are monotonously falling with \ST by construction; however, some of them contain polynomial terms, such that they are not constrained to have a monotonic behavior. In order to determine the background shape, we fit the $N = 3$ \ST distribution or the $N=4$ \ST distribution, normalized to the same total event count as the $N=3$ distribution, in the range of 2.5--4.3\TeV, where any sizable contributions from BSM physics have been ruled out by earlier versions of this analysis, with all 16 functional forms. The lowest masses of the signal models considered, which have not been excluded by the previous analysis~\cite{CMSBH4}, contribute less than 2\% to the total number of events within the fit range. Any functional form observed not to be monotonically decreasing up to $\ST =13\TeV$ after the fit to both multiplicities is discarded. The largest spread among all the accepted functions in the $N = 3$ and $N=4$ fits is used as an envelope of the systematic uncertainty in the background template. The use of both $N=3$ and $N=4$ distributions to construct the envelope allows one to take into account any residual \ST noninvariance in the systematic uncertainty in the background prediction. We observe a good closure of the method to predict the background distributions in simulated QCD multijet events.

\begin{figure}[htb]
    \centering
    \includegraphics[width=0.45\textwidth]{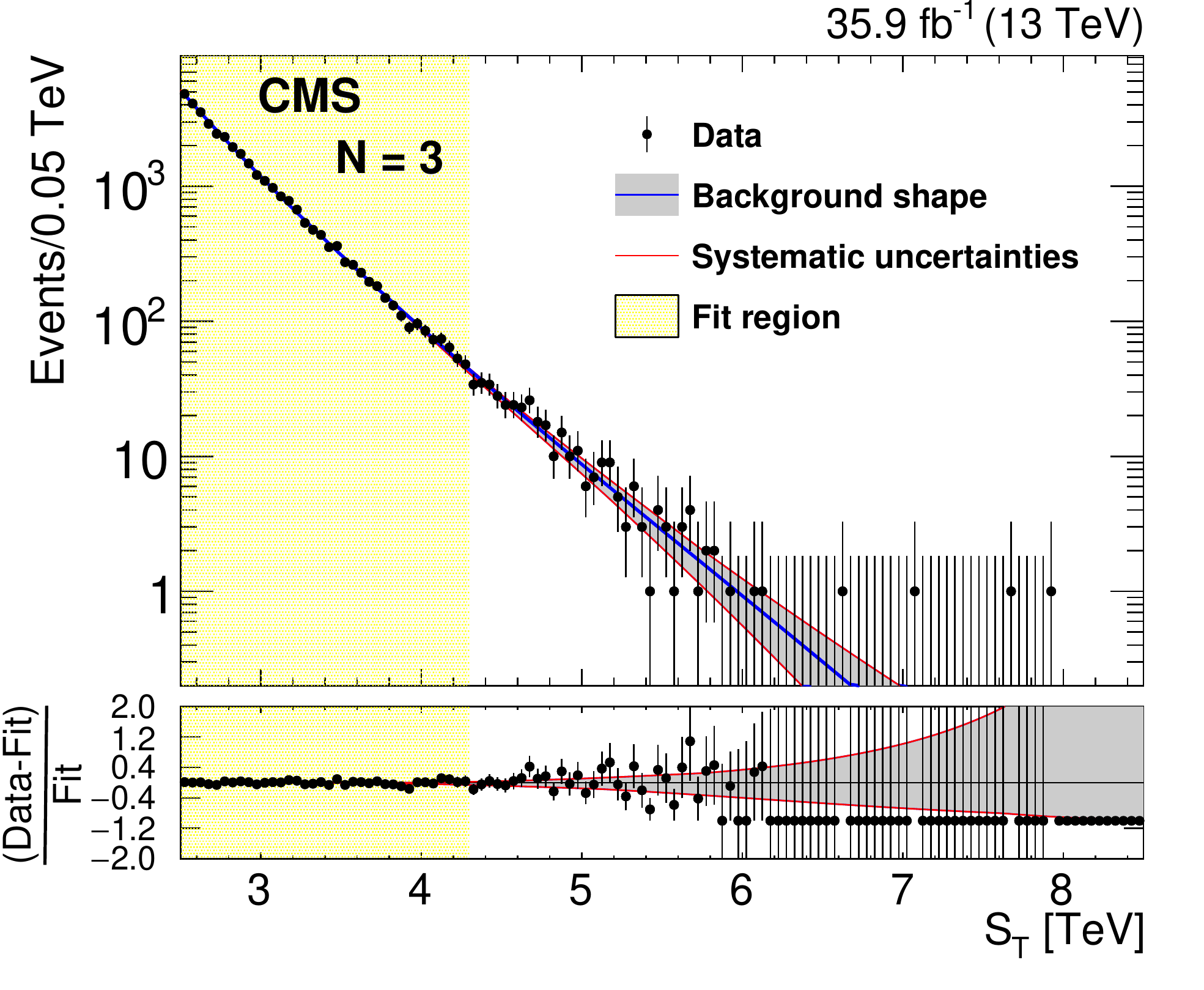}
    \includegraphics[width=0.45\textwidth]{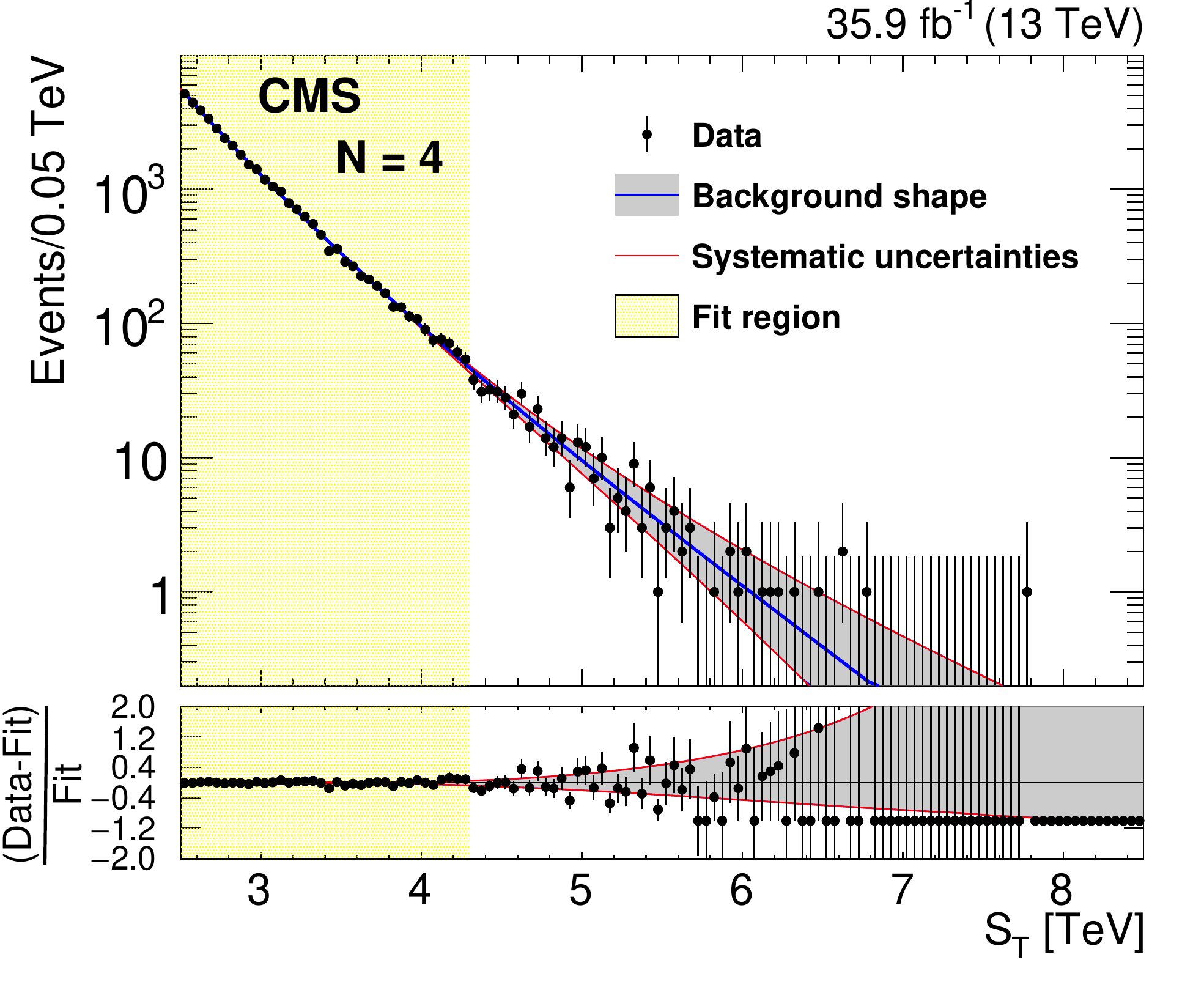}
    \caption{The results of the fit to data with $N = 3$ (left) and $N=4$ (right), after discarding the functions that fail to monotonically
        decrease up to $\ST = 13\TeV$. The description of the best fit function and the envelope are given in the main text. A few points beyond the plotted vertical range in the ratio panels are outside the fit region and do not contribute to the fit quality.}
    \label{fig:fitData}
\end{figure}

The best fits (taking into account the F-test criterion~\cite{F-test} within each set of nested functions) to the $N=3$ and $N=4$ distributions in data, along with the corresponding uncertainty envelopes, are shown in the two panels of Fig.~\ref{fig:fitData}. In both cases, the best fit function is $f(x) = p_0(1-x^{1/3})^{p_1}/(x^{p_2+p_3\log^2(x)})$, where $x = \ST/\sqrt{s} = \ST/(13\TeV)$ and $p_i$ are the four free parameters of the fit. The envelope of the predictions at large \ST ($\ST > 5.5\TeV$, most relevant for the present search) is given by the fit with the following 5-parameter function: $\phi(x) = p_0(1-x)^{p_1}/(x^{p_2 + p_3\log(x)+p_4\log^2(x)})$ to the $N=4$ (upper edge of the envelope) or $N=3$ (lower edge of the envelope) distributions. For \ST values below 5.5\TeV the envelope is built piecewise from other template functions fitted to either the $N=3$ or $N=4$ distribution.

\subsection{Background normalization}

The next step in the background estimation for various inclusive multiplicities is to normalize the template and the uncertainty envelope, obtained as described above, to low-\ST data for various inclusive multiplicities. This has to be done with care, as the \ST invariance is only expected to be observed above a certain threshold, which depends on the inclusive multiplicity requirement. Indeed, since there is a \pt threshold on the objects whose transverse energies count toward the \ST value, the minimum possible \ST value depends on the number of objects in the final state, and therefore the shape invariance for an \ST spectrum with $N \ge \Nmin$ is only observed above a certain \ST threshold, which increases with \Nmin. In order to determine the minimum value of \ST for which this invariance holds, we find a plateau in the ratio of the \ST spectrum for each inclusive multiplicity to that for $N=3$ in simulated multijet events. The plateau for each multiplicity is found by fitting the ratio with a sigmoid function. The lower bound of the normalization region (NR) is chosen to be above the 99\% point of the corresponding sigmoid function. The upper bound of each NR is chosen to be 0.4\TeV above the corresponding lower bound to ensure sufficient event count in the NR. Since the size of the simulated QCD multijet background sample is not sufficient to reliably extract the turn-on threshold for inclusive multiplicities of $N \ge 9$--11, for these multiplicities we use the same NR as for the $N \ge 8$ distribution. A self-consistency check with the CMS data sample has shown that this procedure provides an adequate description of the data. Table \ref{tab:NF} summarizes the turn-on thresholds and the NR boundaries obtained for each inclusive multiplicity.

\begin{table}[htbp!]
    \centering
    \topcaption{The \ST invariance thresholds from fits to simulated QCD multijet background spectra, normalization region definitions, and normalization scale factors in data for different inclusive multiplicities. \label{tab:NF}}
    \begin{tabular}{cccc}
        Multiplicity	 & 99\% turn-on & Normalization & Normalization  \\
        			 & point (\TeVns{}) & region (\TeVns{})& scale factor (data) \\
        \hline
        ${\ge}3$ 	& $2.44 \pm 0.06$ & 2.5--2.9 & $3.437 \pm 0.025$  \\
        ${\ge}4$ 	& $2.47 \pm 0.06$ & 2.5--2.9 & $2.437 \pm 0.019$  \\
        ${\ge}5$ 	& $2.60 \pm 0.07$ & 2.7--3.1 & $1.379 \pm 0.016$  \\
        ${\ge}6$ 	& $2.75 \pm 0.11$ & 2.9--3.3 & $0.652 \pm 0.012$  \\
        ${\ge}7$ 	& $2.98 \pm 0.13$ & 3.0--3.4 & $0.516 \pm 0.015$  \\
        ${\ge}8$ 	& $3.18 \pm 0.21$ & 3.2--3.6 & $0.186 \pm 0.011$  \\
        ${\ge}9$ 	& $3.25 \pm 0.28$ & 3.2--3.6 & $0.055 \pm 0.006$  \\
        ${\ge}10$ 	& $3.02 \pm 0.26$ & 3.2--3.6 & $0.012 \pm 0.003$  \\
        ${\ge}11$ 	& $2.89 \pm 0.24$ & 3.2--3.6 & $0.002 \pm 0.001$  \\
    \end{tabular}
   \end{table}
\begin{figure}[htbp!]
    \centering
    \includegraphics[width=0.47\textwidth]{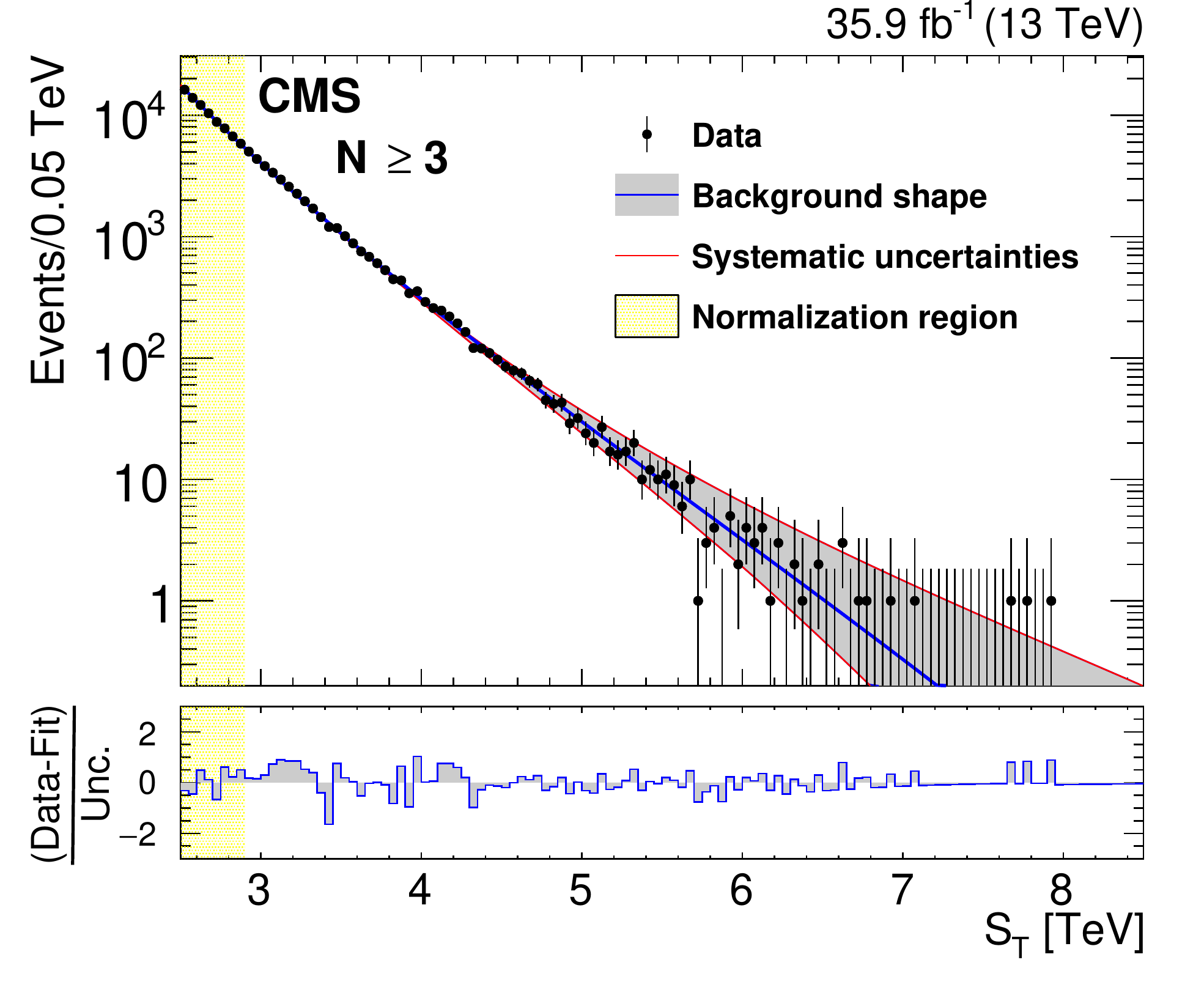}
    \includegraphics[width=0.47\textwidth]{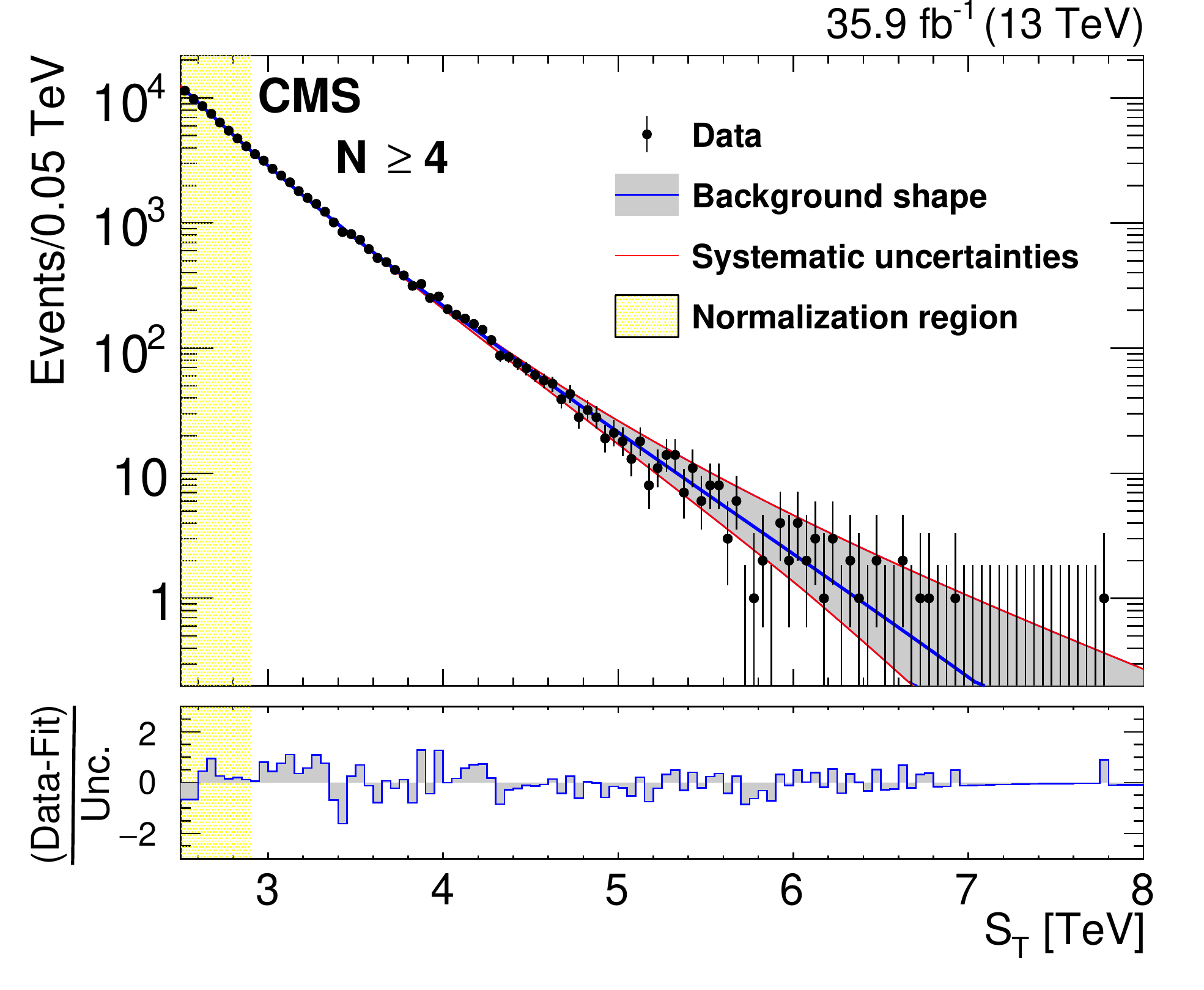}
    \includegraphics[width=0.47\textwidth]{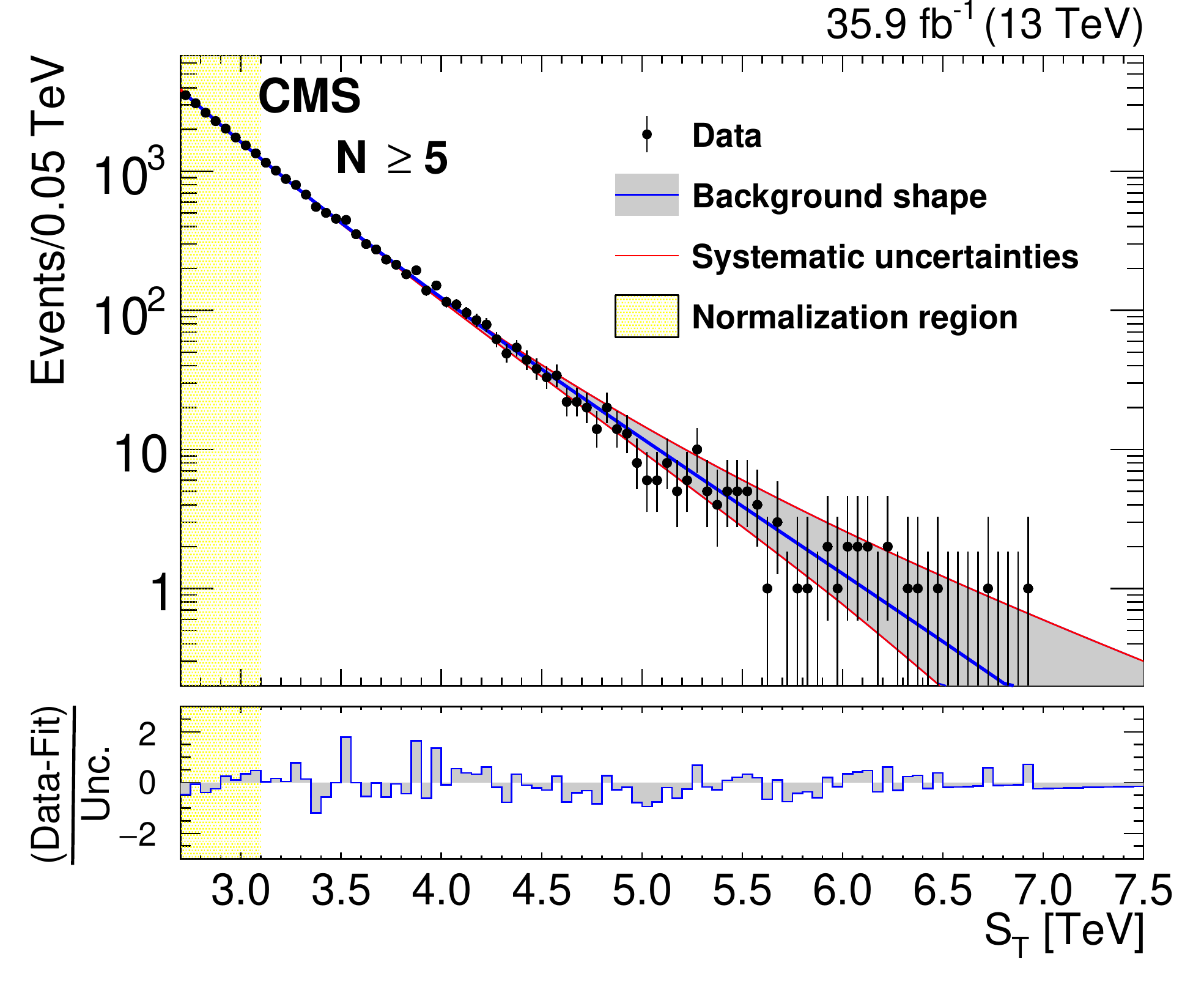}
    \includegraphics[width=0.47\textwidth]{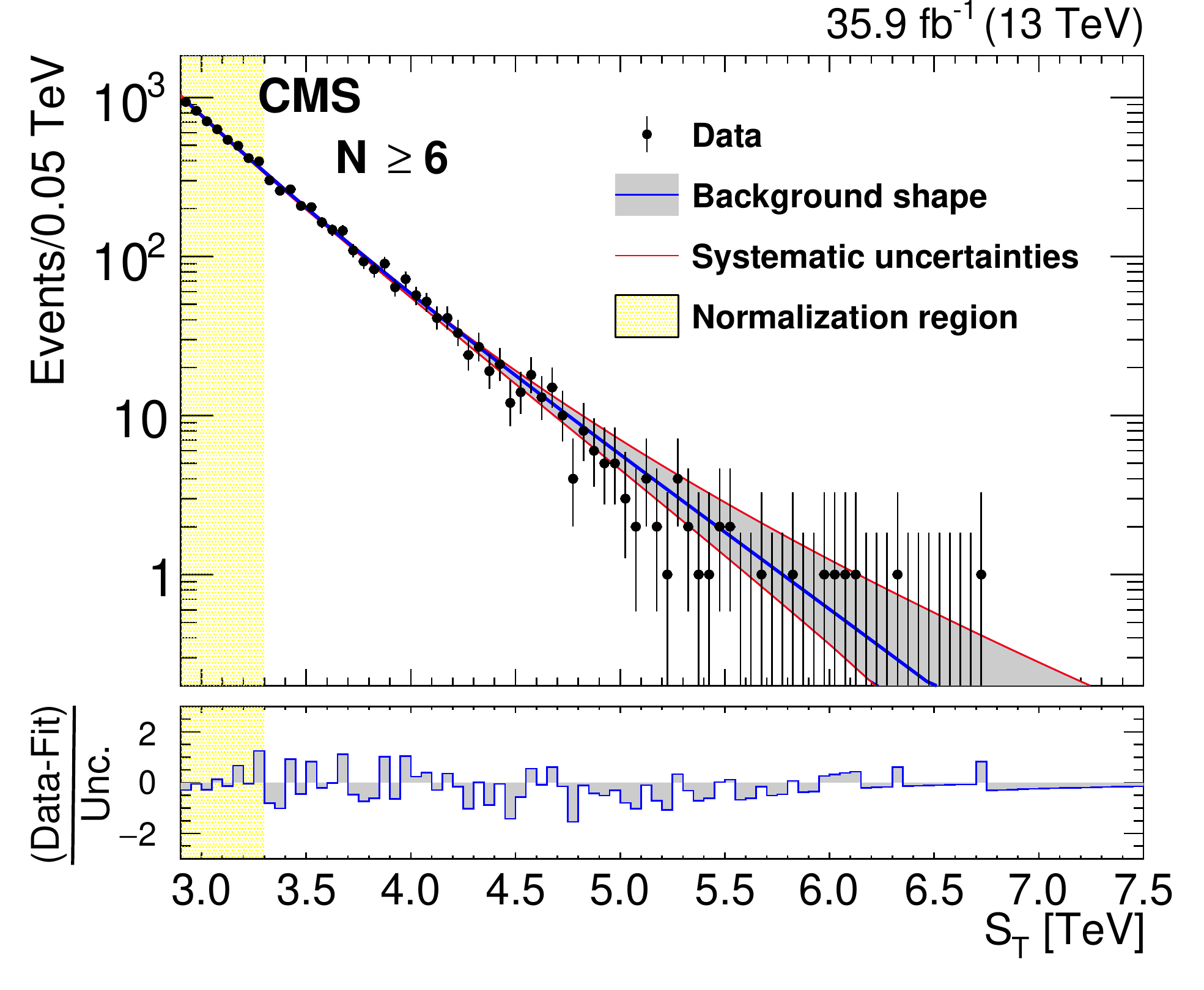}
    \caption{The comparison of data and the background predictions after the normalization for inclusive multiplicities $N \ge 3, \dots, 6$ (left to right, upper to lower).
        The gray band shows the background shape uncertainty alone and the red lines also include the normalization uncertainty.
        The bottom panels show the difference between the data and the background prediction from the fit, divided by the overall uncertainty, which includes the statistical uncertainty of data as well as the shape and normalization uncertainties in the background prediction, added in quadrature.}
    \label{fig:dataBkg3to6}
\end{figure}

The normalization scale factors  are calculated as the ratio of the number of events in each NR for the inclusive multiplicities of $N\geq 3, \dots, 11$ to that for the exclusive multiplicity of $N = 3$ in data, and are listed in Table \ref{tab:NF}.
The relative scale factor uncertainties are derived from the number of events in each NR, as $1/\sqrt{N_\mathrm{NR}}$, where $N_\mathrm{NR}$ is the number of events in the corresponding NR.

\subsection{Comparison with data}

\begin{figure}[htbp!]
    \centering
    \includegraphics[width=0.47\textwidth]{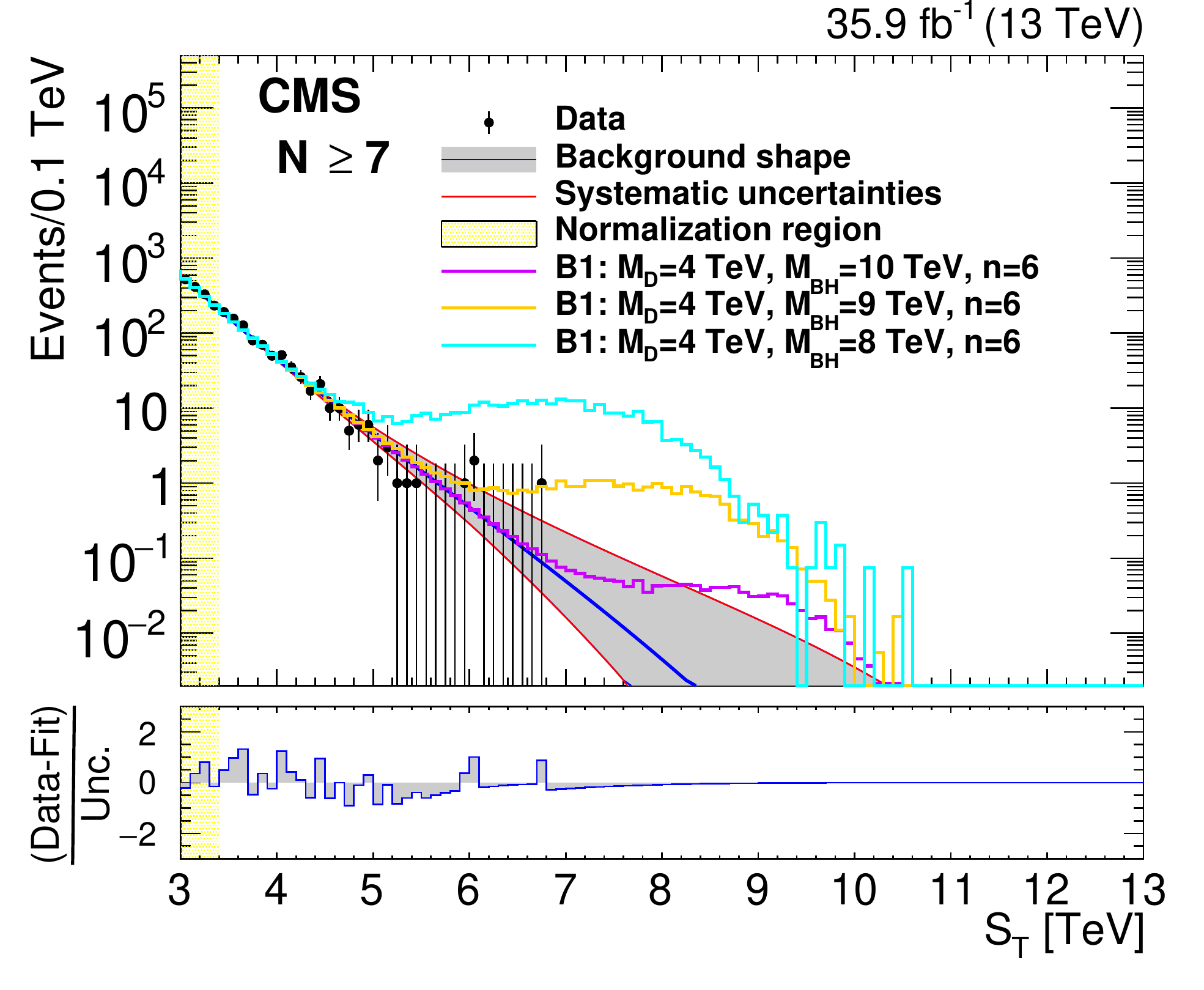}
    \includegraphics[width=0.47\textwidth]{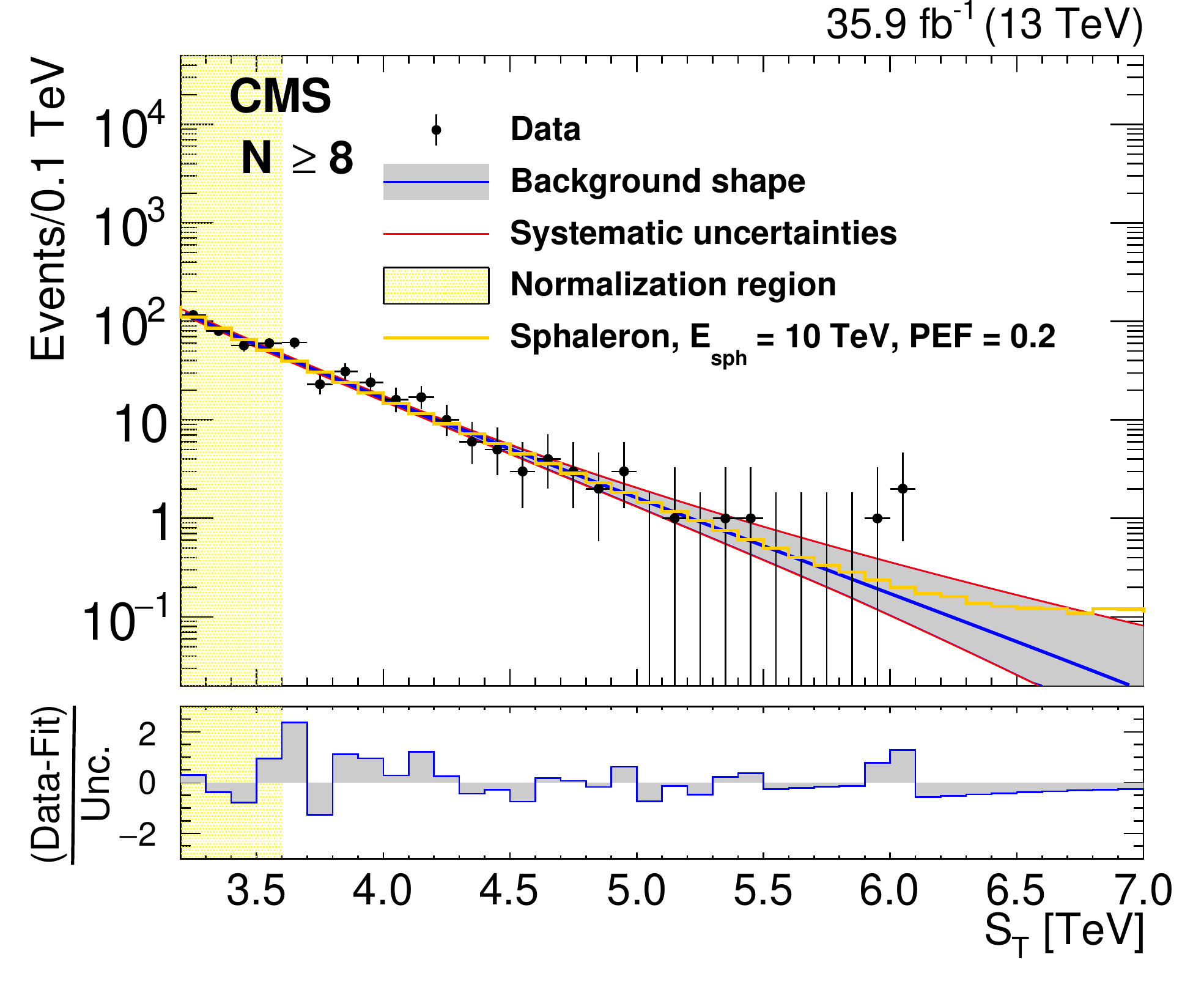}
    \includegraphics[width=0.47\textwidth]{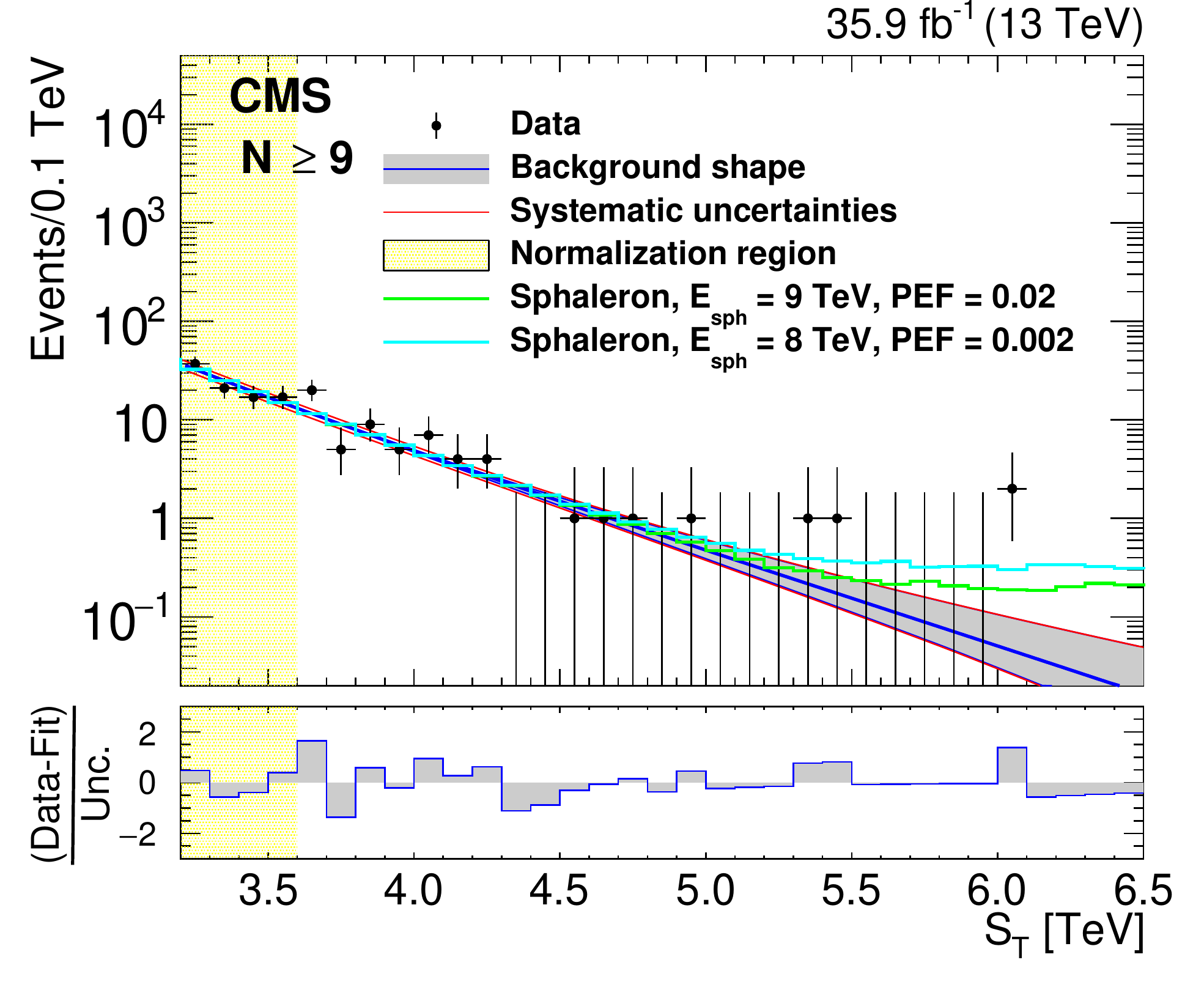}
    \includegraphics[width=0.47\textwidth]{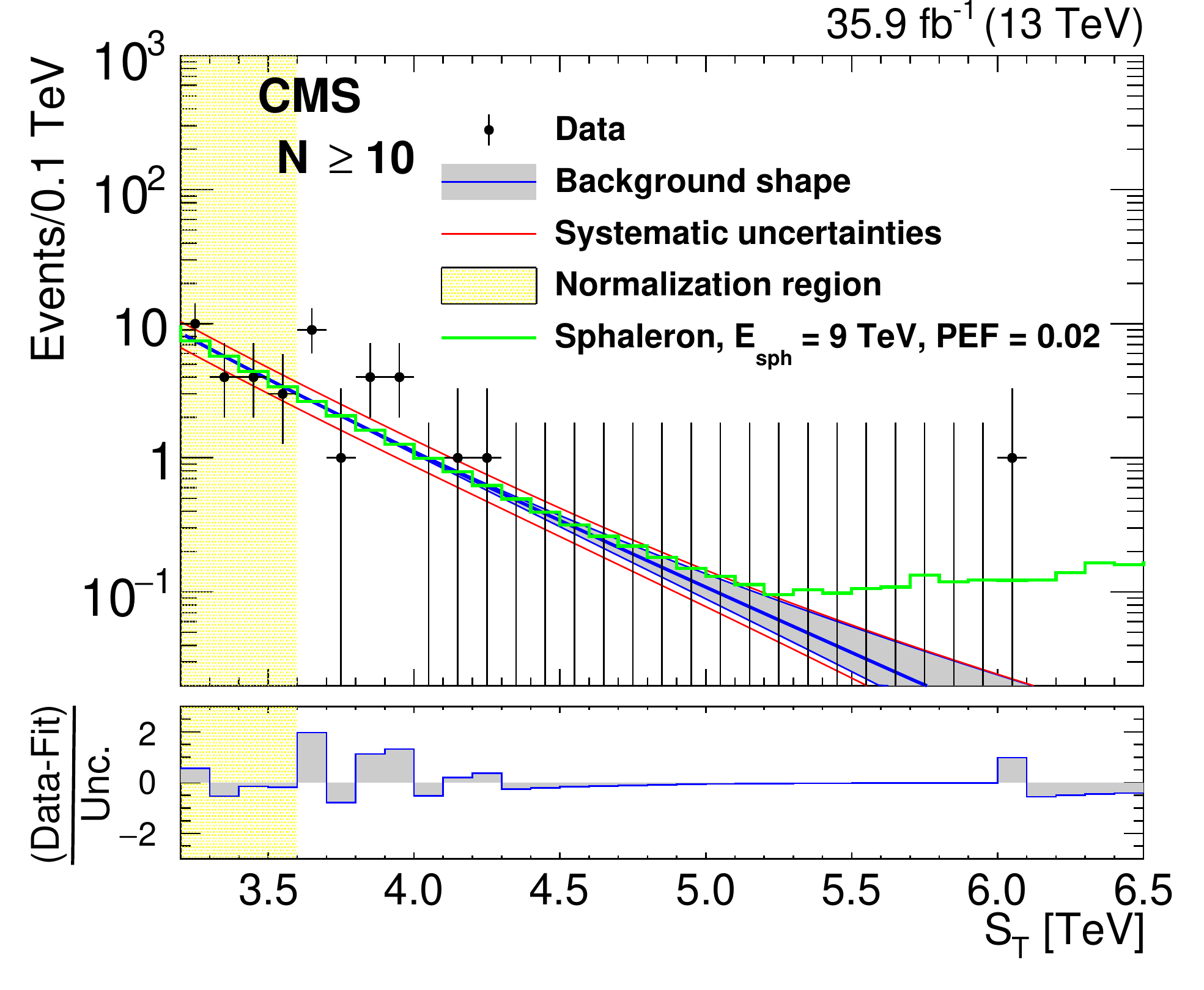}
    \includegraphics[width=0.47\textwidth]{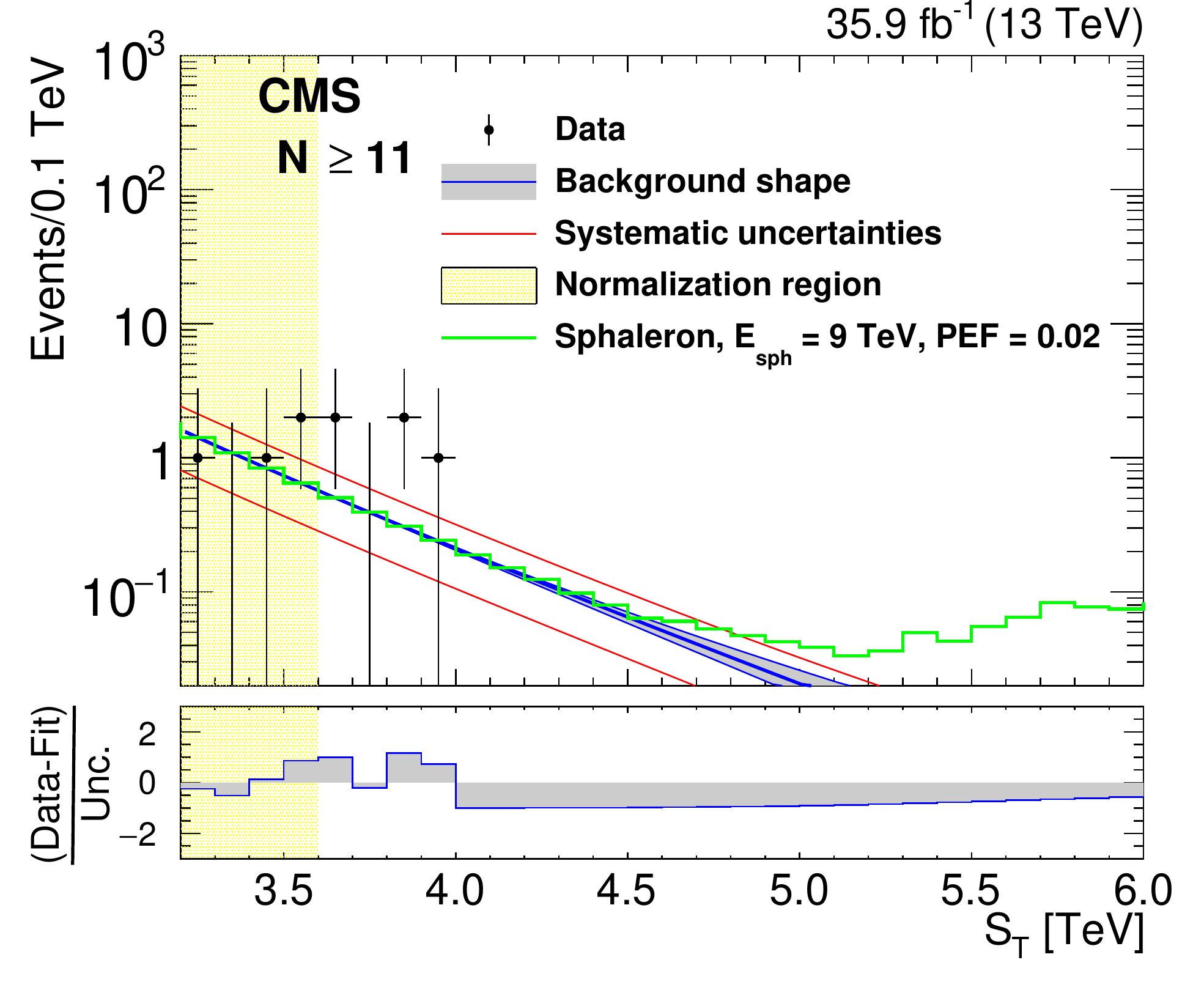}
    \caption{The comparison of data and the background predictions after normalization for inclusive multiplicities of $ N \ge 7, \dots, 11$ (left to right, upper to lower).
        The gray band shows the shape uncertainty and the red lines also include the normalization uncertainty.
        The bottom panels show the difference between the data and the background prediction from the fit, divided by the overall uncertainty, which includes the statistical uncertainty of data as well as the shape and normalization uncertainties in the background prediction, added in quadrature. The $N \ge 7$ ($N \ge 8, \dots, 11$) distributions also show contributions from benchmark \BLACKMAX B1 (sphaleron) signals added to the expected background.}
    \label{fig:dataBkg7to11}
\end{figure}

The results of the background prediction and their comparison with the observed data are shown in Figs.~\ref{fig:dataBkg3to6} and \ref{fig:dataBkg7to11} for inclusive multiplicities $N \ge 3,\dots, 11$. The data are consistent with the background predictions in the entire \ST range probed, for all inclusive multiplicities.

\section{Systematic uncertainties\label{s:systematics}}

There are several sources of systematic uncertainty in this analysis. Since the background estimation is based on control samples in data, the only uncertainties affecting the background predictions are the modeling of the background shape via template functions and the normalization of the chosen function to data at low \ST, as described in Section~\ref{s:backgrounds}. They are found to be 1--130\% and  0.7--50\%, depending on the values of \ST and \Nmin, respectively.

For the signal, we consider the uncertainties in the PDFs, jet energy scale (JES), and the integrated luminosity. For the PDF uncertainty, we only consider the effect on the signal acceptance, while the PDF uncertainty in the signal cross section is treated as a part of the theoretical uncertainty and therefore is not propagated in the experimental cross section limit. The uncertainty in the signal acceptance is calculated using PDF4LHC recommendations~\cite{PDF4LHC1,PDF4LHC2} based on the quadratic sum of variations from the MSTW2008 uncertainty set (${\approx}0.5\%$), as well as the variations obtained by using three different PDF sets: MSTW2008, CTEQ6.1~\cite{CTEQ}, and NNPDF2.3~\cite{NNPDF} (up to 6\% based on the difference between the default and CTEQ6.1 sets) for one of the benchmark models (nonrotating BH with $\MD= 3\TeV$, $\MBH = 5.5\TeV$, and $n = 2$, as generated by \BLACKMAX); the size of the effect for other benchmark points is similar. To be conservative, we assign a systematic uncertainty of 6\%
due to the choice of PDFs for all signal samples. The JES uncertainty affects the signal acceptance because of the kinematic requirements on the objects and the fraction of signal events passing a certain \STmin threshold used for limit setting, as described in Section~\ref{s:Limits}. In order to account for these effects, the jet four-momenta are simultaneously shifted up or down by the JES uncertainty, which is a function of the jet \pt and
$\eta$, and the largest of the two differences with respect to the use of the nominal JES is assigned as the uncertainty. The uncertainty due to JES depends on \MBH and varies between $<$1 and 5\%; we conservatively assign a constant value of 5\% as the signal acceptance uncertainty due to JES. Finally, the integrated luminosity is measured with an uncertainty of 2.5\%~\cite{LUM-17-001}. Effects of all other uncertainties on the signal acceptance are negligible.

The values of systematic uncertainties that are used in this analysis are summarized in Table~\ref{tab:Table_Uncertainties}.

\begin{table}[htbp!]
\centering
    \topcaption{Summary of systematic uncertainties in the signal acceptance and the background estimate.\label{tab:Table_Uncertainties}}
    \begin{tabular}{lcc}
       	Uncertainty source		& Effect on signal acceptance 	& Effect on background\\
	\hline
   	PDF 			        & $\pm 6\%$ 			        & \NA \\
   	JES 	                & $\pm 5\%$ 			        & \NA \\
   	Integrated luminosity 	& $\pm 2.5\%$ 			        & \NA \\
   	Shape modeling 		    & \NA 				            & $\pm$(1--130)\%, depending on \ST\\
   	Normalization 		    & \NA 				            & $\pm$(0.7--50)\%, depending on \Nmin \\
    \end{tabular}
\end{table}

\section{Results\label{s:Limits}}

As shown in Figs.~\ref{fig:dataBkg3to6} and \ref{fig:dataBkg7to11}, there is no evidence for a statistically significant signal observed in any of the inclusive \ST distributions. The null results of the search are interpreted in terms of model-independent limits on BSM physics in energetic, multiparticle final states, and as model-specific limits for a set of semiclassical BH and SB scenarios, as well as for EW sphalerons.

Limits are set using the \CLs method~\cite{Junk,Read,ATLAS_CMS} with log-normal priors in the likelihood to constrain the nuisance parameters near their best estimated values. We do not use an asymptotic approximation of the \CLs method~\cite{Gross}, as for most of the models the optimal search region corresponds to a very low background expectation, in which case the asymptotic approximation is known to overestimate the search sensitivity.

\subsection{Model-independent limits}

\begin{figure}[htpb!]
    \centering
    \includegraphics[width=0.45\textwidth]{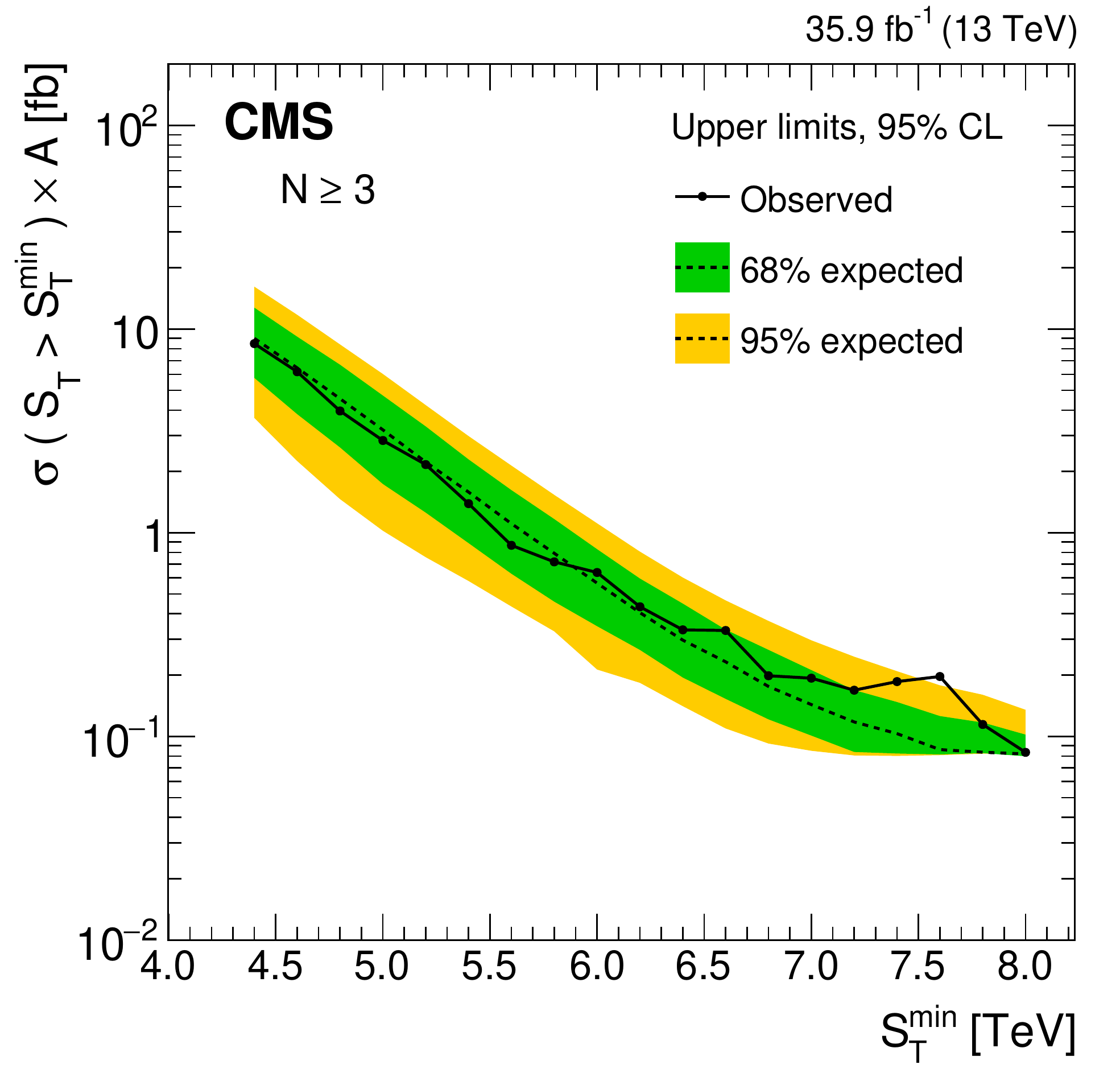}
    \includegraphics[width=0.45\textwidth]{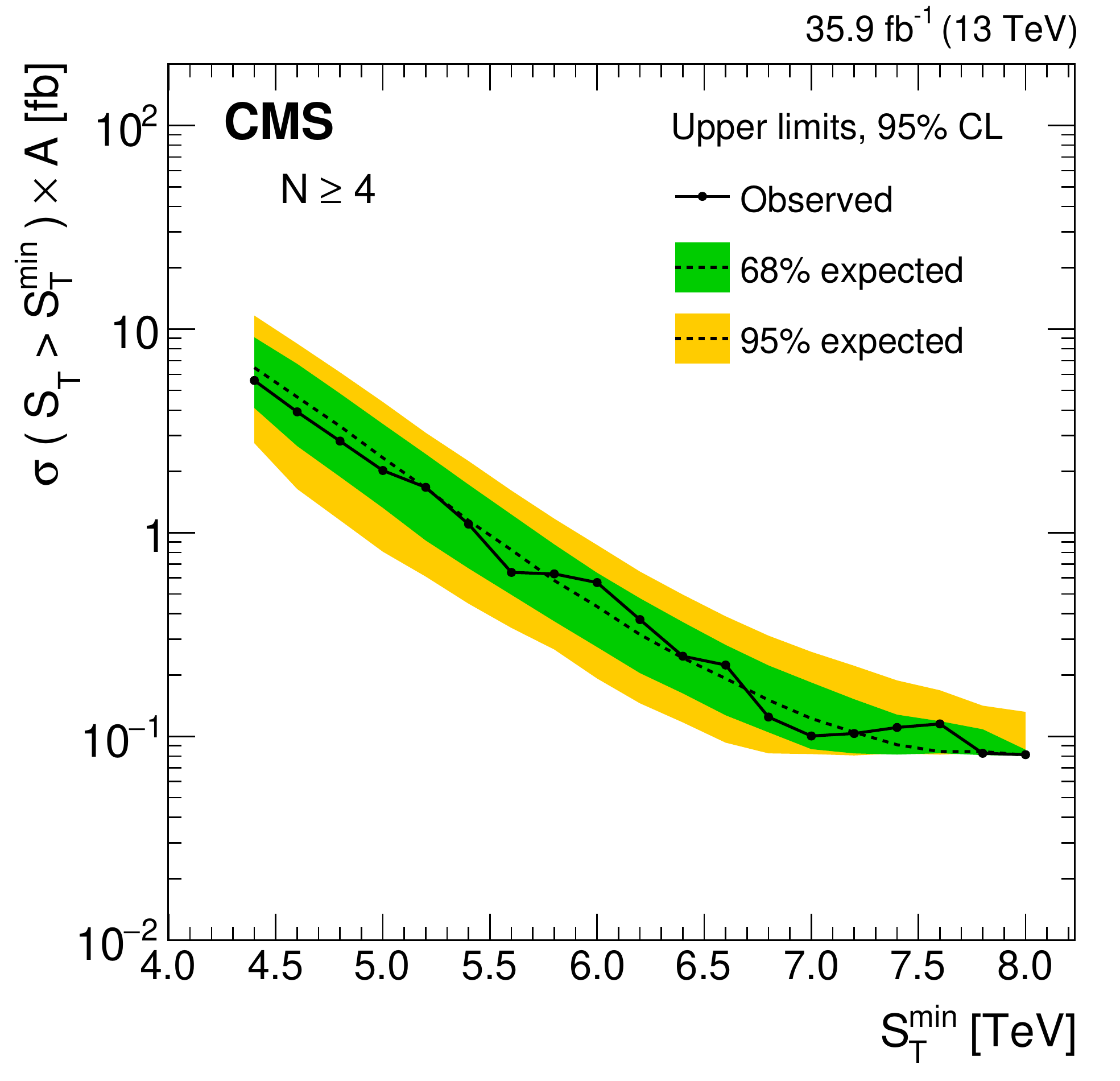}
    \includegraphics[width=0.45\textwidth]{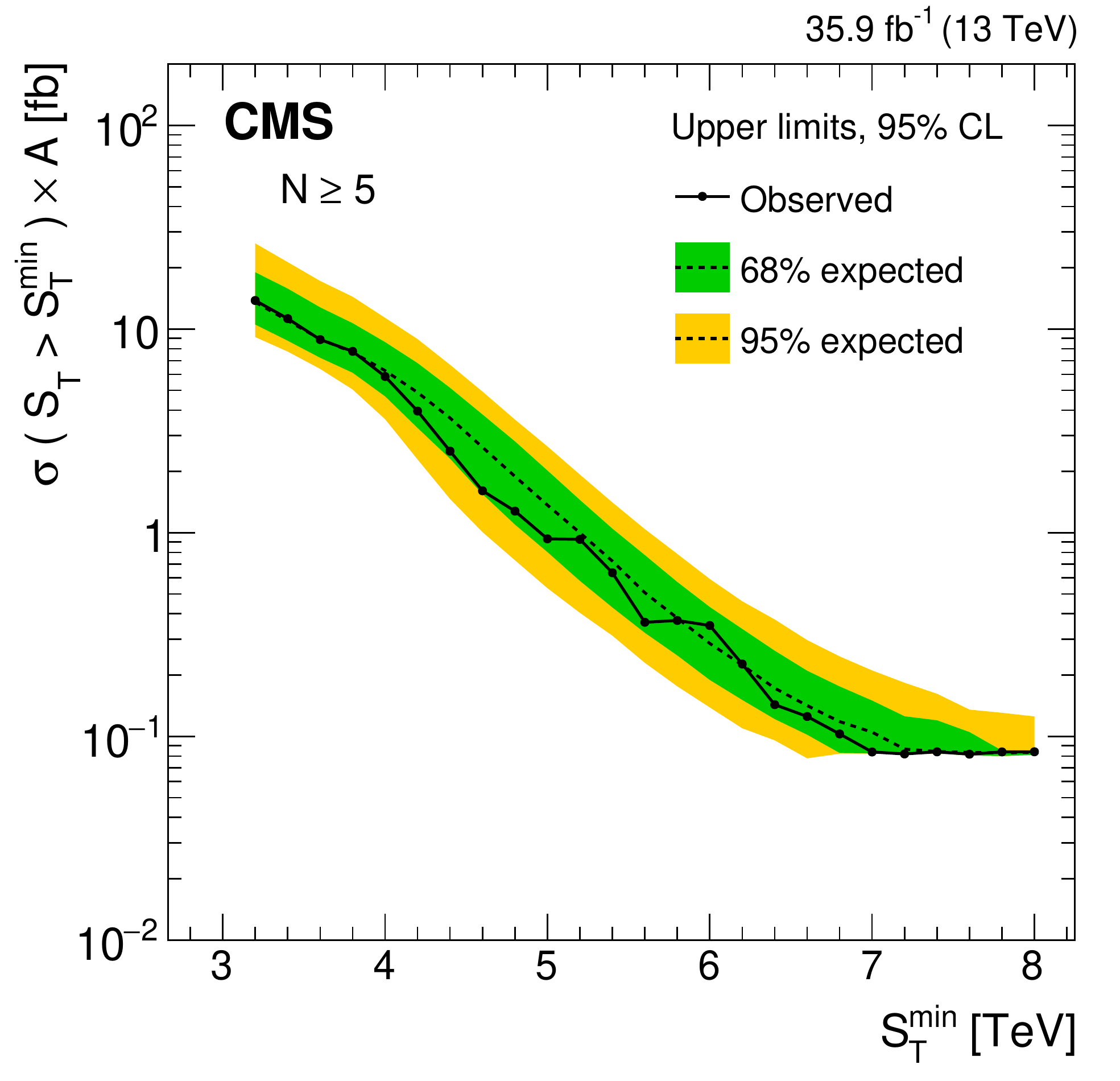}
    \includegraphics[width=0.45\textwidth]{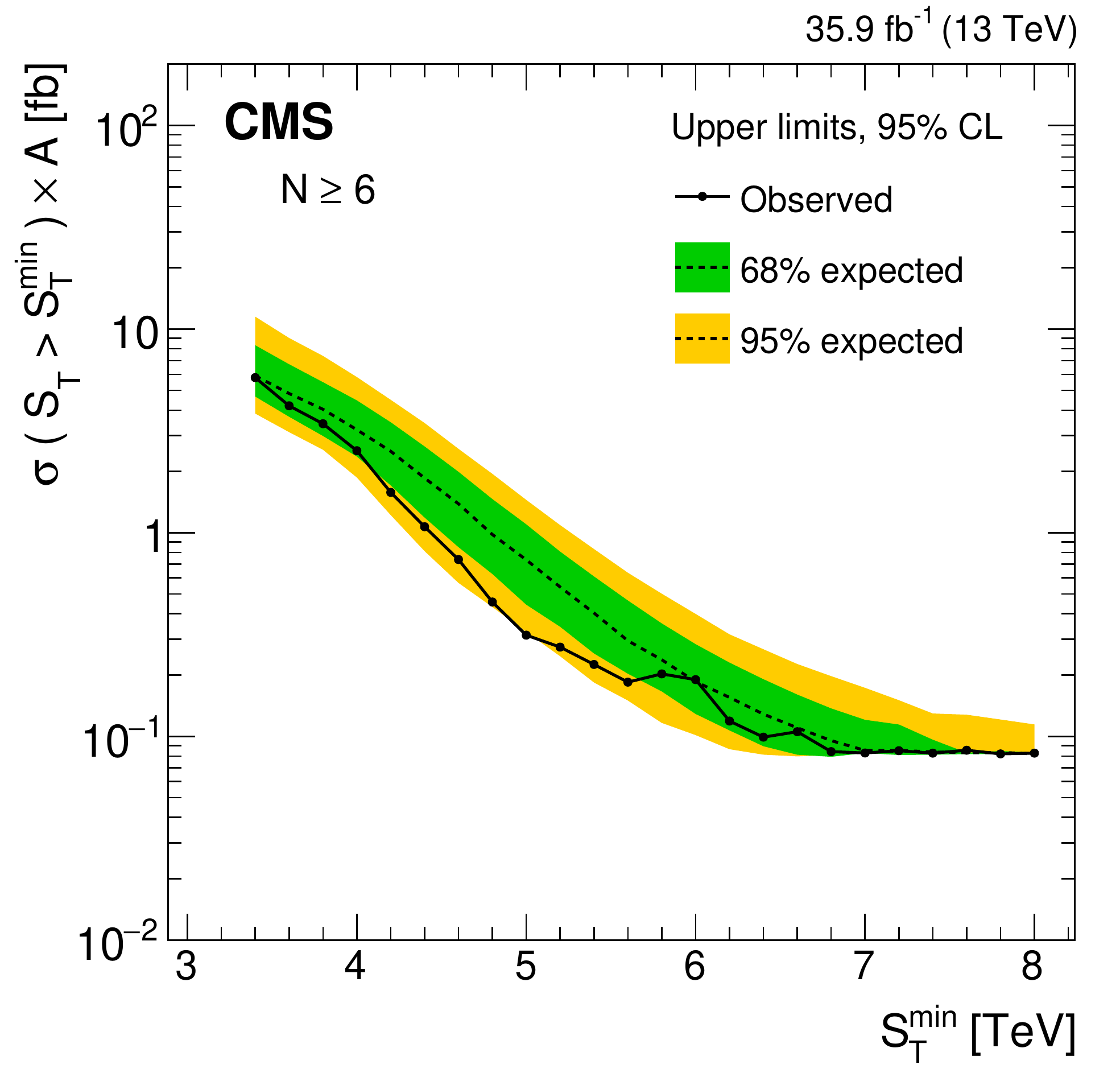}
    \caption{Model-independent upper limits on the cross section times acceptance for four sets of inclusive multiplicity thresholds, $N \ge 3, \dots, 6$ (left to right, upper to lower). Observed (expected) limits are shown as the black solid (dotted) lines. The inner (outer) band represents the $\pm 1$ ($\pm 2$) standard deviation uncertainty in the expected limit.}
    \label{fig:Limit_ModelIndependent1}
\end{figure}
\begin{figure}[htbp!]
    \centering
    \includegraphics[width=0.45\textwidth]{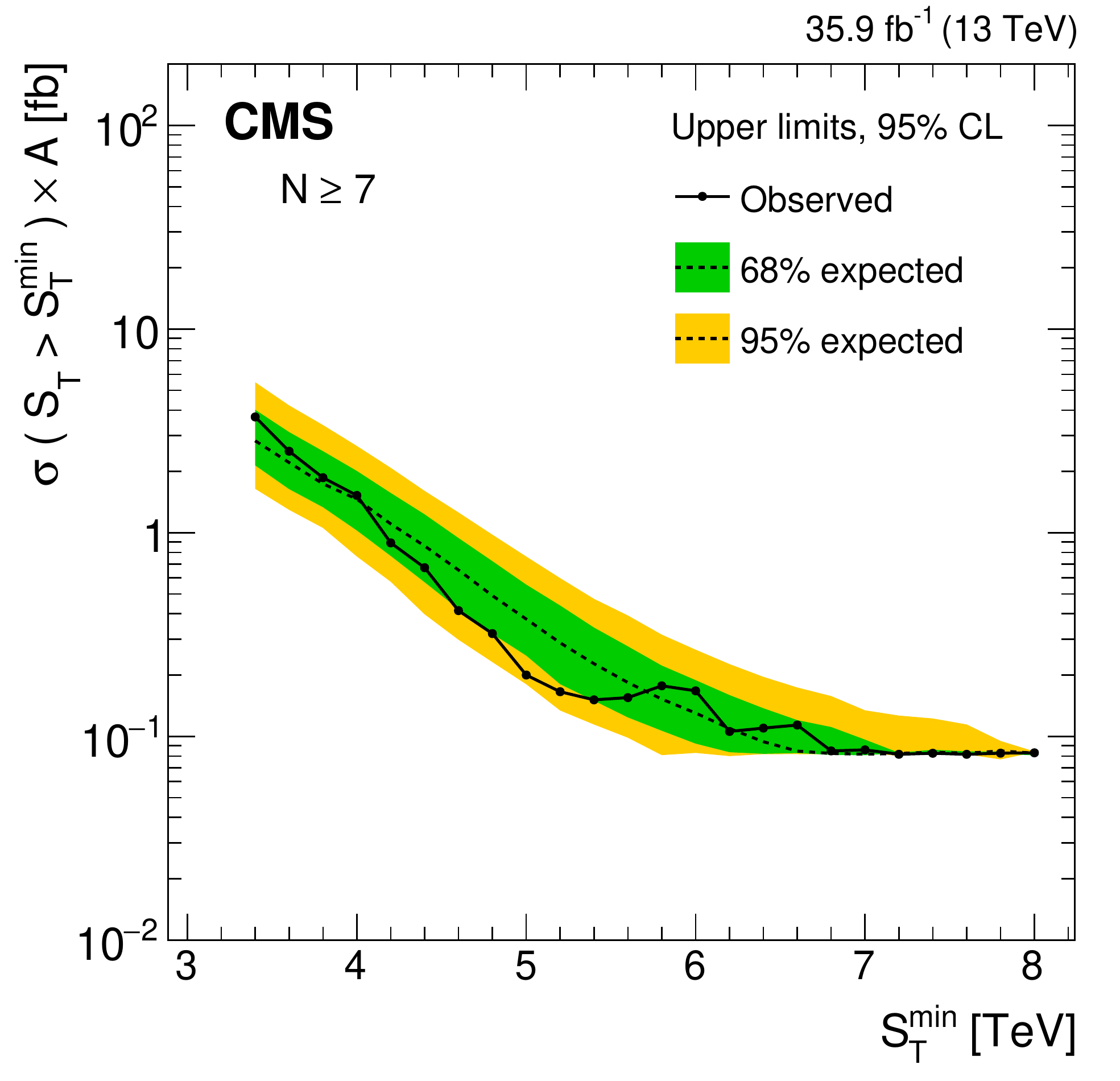}
    \includegraphics[width=0.45\textwidth]{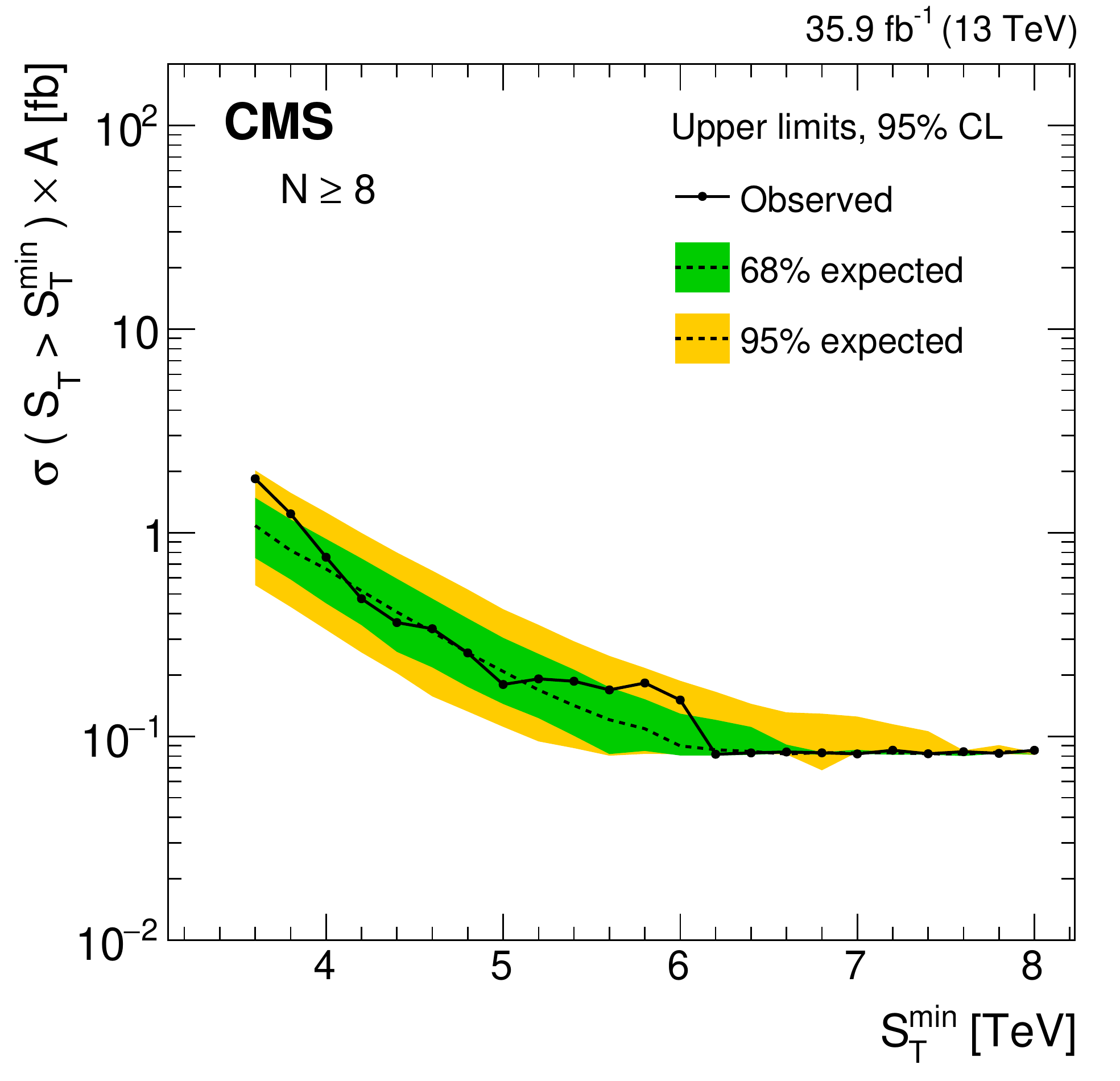}
    \includegraphics[width=0.45\textwidth]{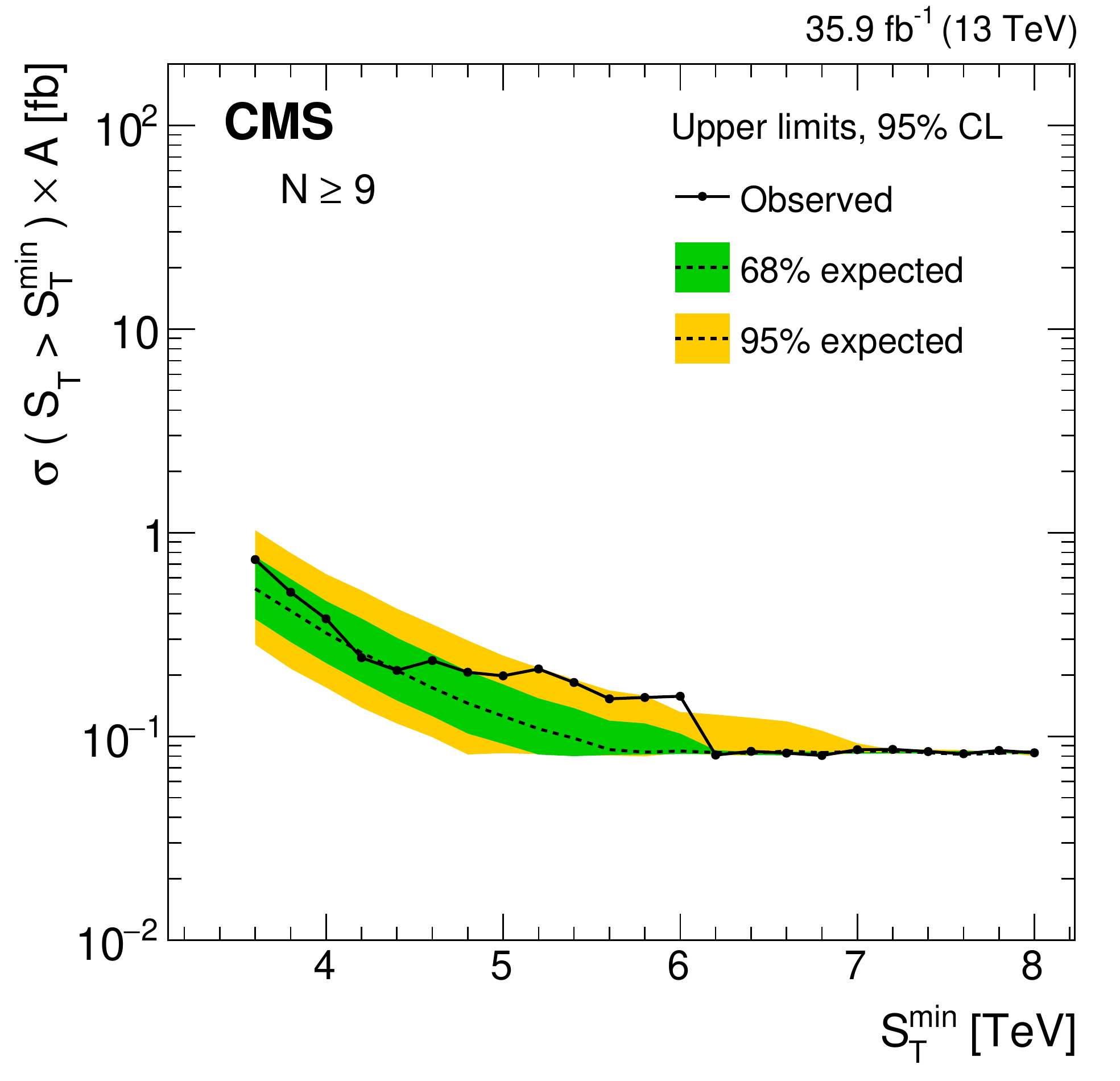}
    \includegraphics[width=0.45\textwidth]{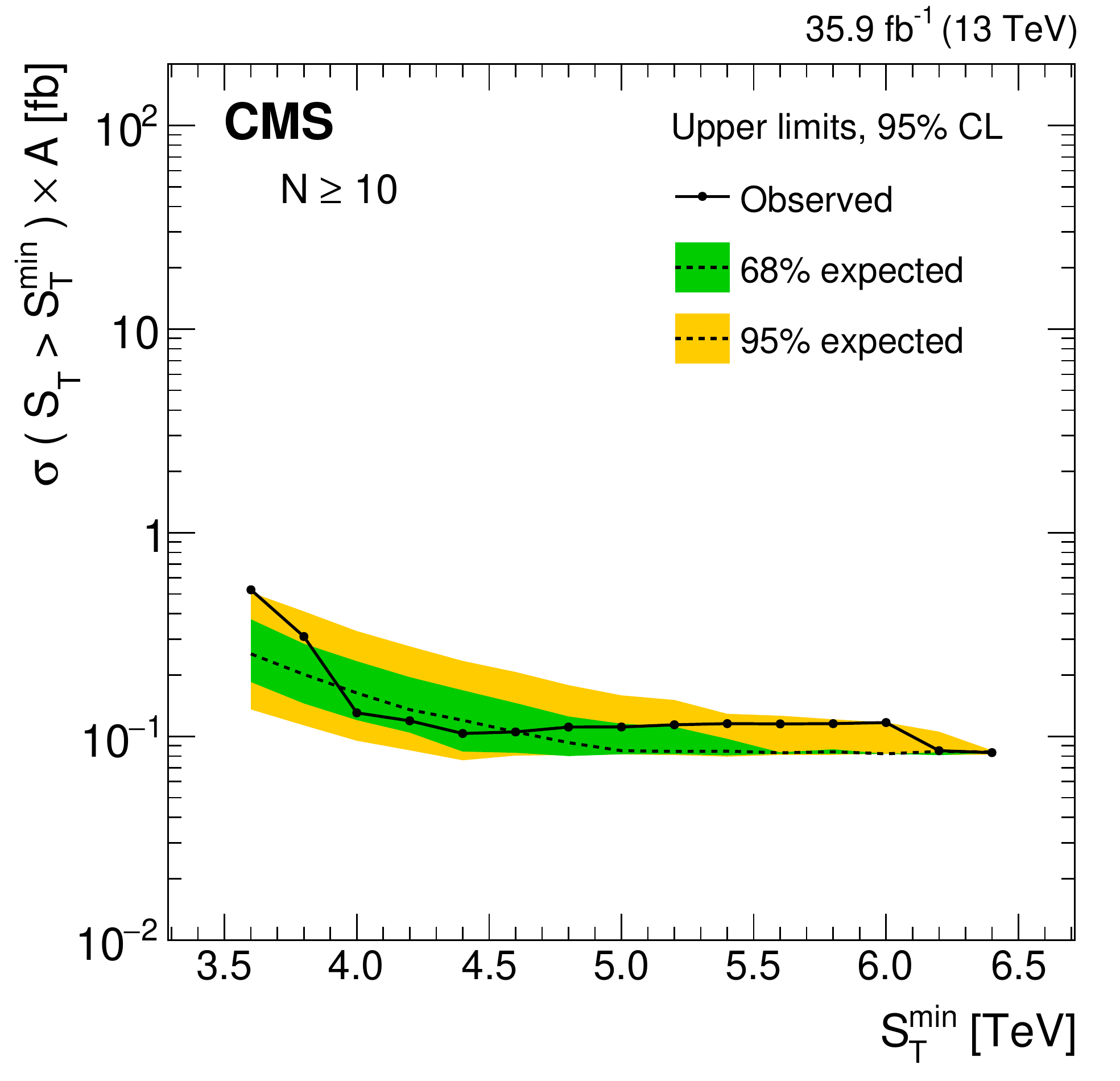}
    \includegraphics[width=0.45\textwidth]{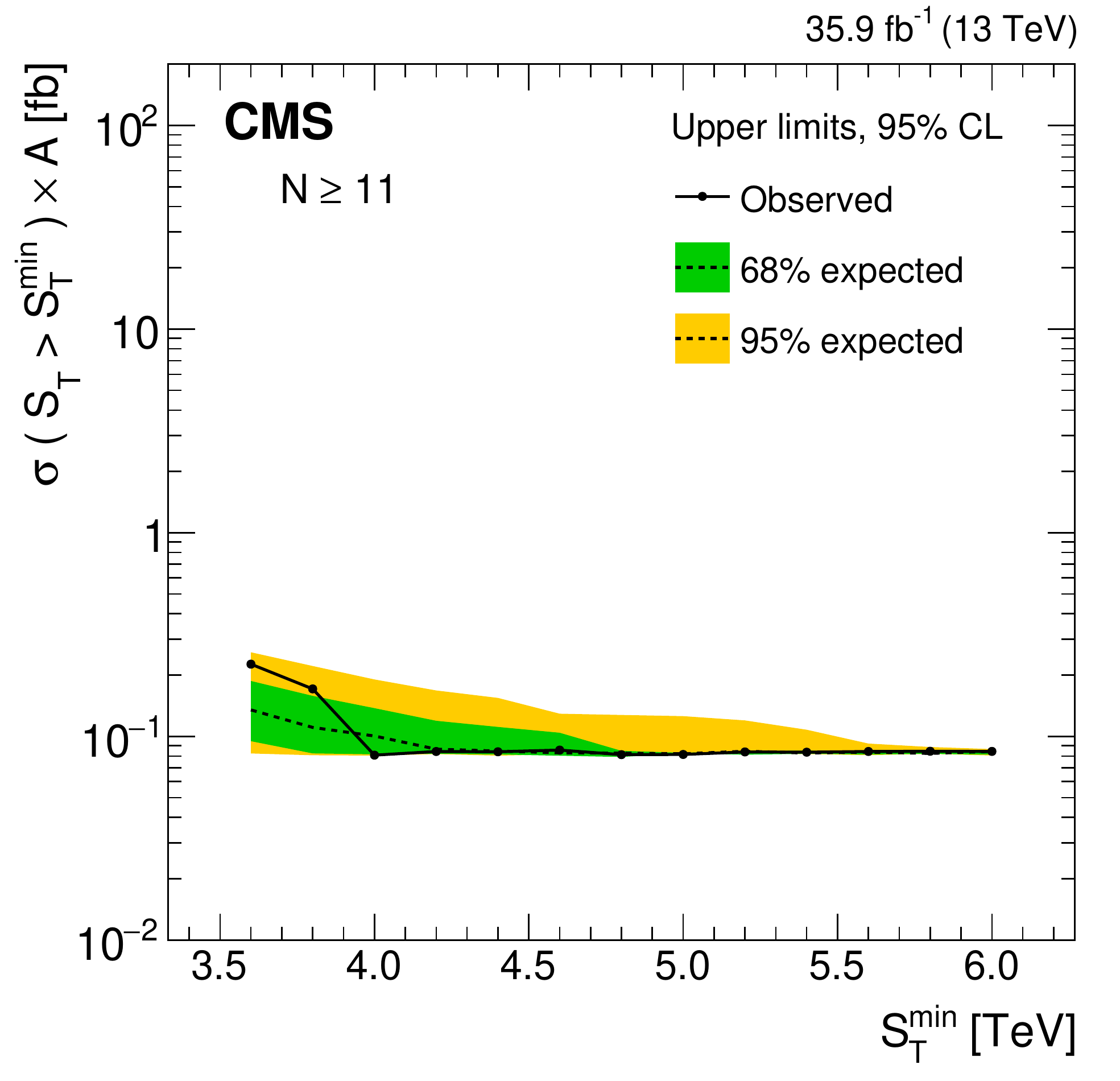}
    \caption{Model-independent upper limits on the cross section times acceptance for five sets of
        inclusive multiplicity thresholds, $N \ge 7, \dots, 11$ (left to right, upper to lower).
        Observed (expected) limits are shown as the black solid (dotted) lines. The inner (outer) band represents the $\pm 1$ ($\pm 2$) standard deviation uncertainty in the expected limit.}
    \label{fig:Limit_ModelIndependent2}
\end{figure}

The main result of this analysis is a set of model-independent upper limits on the product of signal cross section and acceptance ($\sigma \, A$)
in inclusive $N \ge \Nmin$ final states, as a function of the minimum \ST requirement, \STmin, obtained from a simple counting experiment
for $\ST > \STmin$.
These limits can then be translated into limits on the \MBHmin in a variety of models, or on any
other signals resulting in an energetic, multi-object final state. We start with the limits for the inclusive multiplicities
$N \ge 3,4$, which can be used to constrain
models resulting in lower multiplicities of the final-state objects. Since part of the data entering these distributions are used to determine the background shape and its uncertainties, the limits are set only for \STmin values above the background fit region, \ie, for $\ST > 4.5\TeV$. For other multiplicities, the limits are shown for \ST values above the NRs listed in Table~\ref{tab:NF}.
These limits at 95\% confidence level (\CL) are shown in Figs.~\ref{fig:Limit_ModelIndependent1} and~\ref{fig:Limit_ModelIndependent2}. When computing the limits, we use systematic uncertainties in the signal acceptance applicable to the specific models discussed in this paper, as documented
in Section~\ref{s:systematics}. It is reasonable to expect these limits to apply to a large variety of models resulting in multi-object final states dominated by jets.
The limits on the product of the cross section and acceptance approach 0.08\unit{fb} at high values of \STmin.

\subsection{Model-specific limits}

To determine the optimal point of \STmin and the minimum multiplicity of the final-state objects \Nmin for setting an exclusion limit for a particular model, we calculate the acceptance and the expected limit on the cross section for a given model for each point of the model-independent limit curves, for all inclusive multiplicities. The optimal point of $(\Nmin,\STmin)$ is chosen as the point that gives the lowest expected cross section limit. In most of the cases this point also maximizes the significance of an observation, for the case of a nonzero signal present in data~\cite{CMSBH4}.

\begin{figure}[htb]
    \centering
    \includegraphics[height=0.5\textwidth]{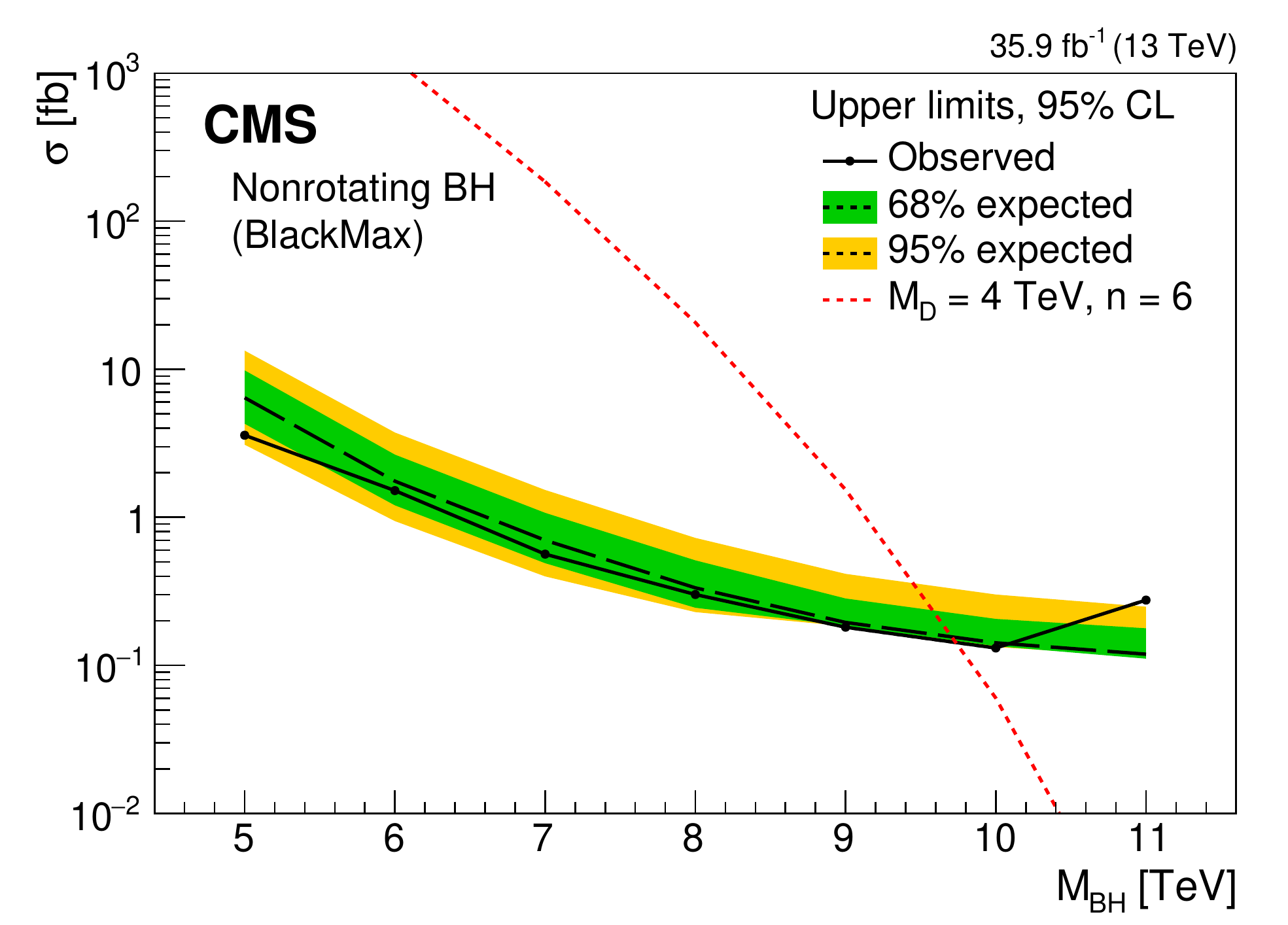}
    \caption{Example of a model-specific limit on \MBHmin for a semiclassical nonrotating BH model (\BLACKMAX point B1) with $\MD = 4\TeV$ $\nED=6$, as a function of \MBHmin. The 95\% \CL upper exclusion limit on the signal cross section for each \MBHmin value is obtained at the optimal $(\Nmin,\STmin)$ point, which ranges from $(7,5.0\TeV)$ for $\MBHmin = 5\TeV$ to $(3,7.6\TeV)$ for $\MBHmin = 11\TeV$.
Also shown with a dashed line are the theoretical cross sections corresponding to these optimal points. The inner (outer) band represents the $\pm 1$ ($\pm 2$) standard deviation uncertainty in the expected limit.}
    \label{fig:BH_optPoint}
\end{figure}

An example of a model-specific limit is given in Fig.~\ref{fig:BH_optPoint} for a \BLACKMAX benchmark point B1 (nonrotating semiclassical BH) with \MD = 4\TeV, $\nED =6$, and \MBHmin between 5 and 11\TeV. In this case, the optimal inclusive multiplicity \Nmin starts at 7 for the lowest \MBHmin value of 5\TeV, with the corresponding $\STmin = 5\TeV$. As \MBHmin increases, the optimal point shifts to lower inclusive multiplicities and the corresponding \STmin increases, reaching $(3,7.6\TeV)$ for $\MBHmin = 11\TeV$. The corresponding 95\% \CL upper limit curve and the theoretical cross section for the chosen benchmark point is shown in Fig.~\ref{fig:BH_optPoint}. The observed (expected) 95\% \CL lower limit on \MBHmin in this benchmark model can be read from this plot as the intersection of the theoretical curve with the observed (expected) 95\% \CL upper limit on the cross section, and is found to be 9.7 (9.7)\TeV.

\begin{figure}[htb!]
    \centering
    \includegraphics[height=0.6\textwidth]{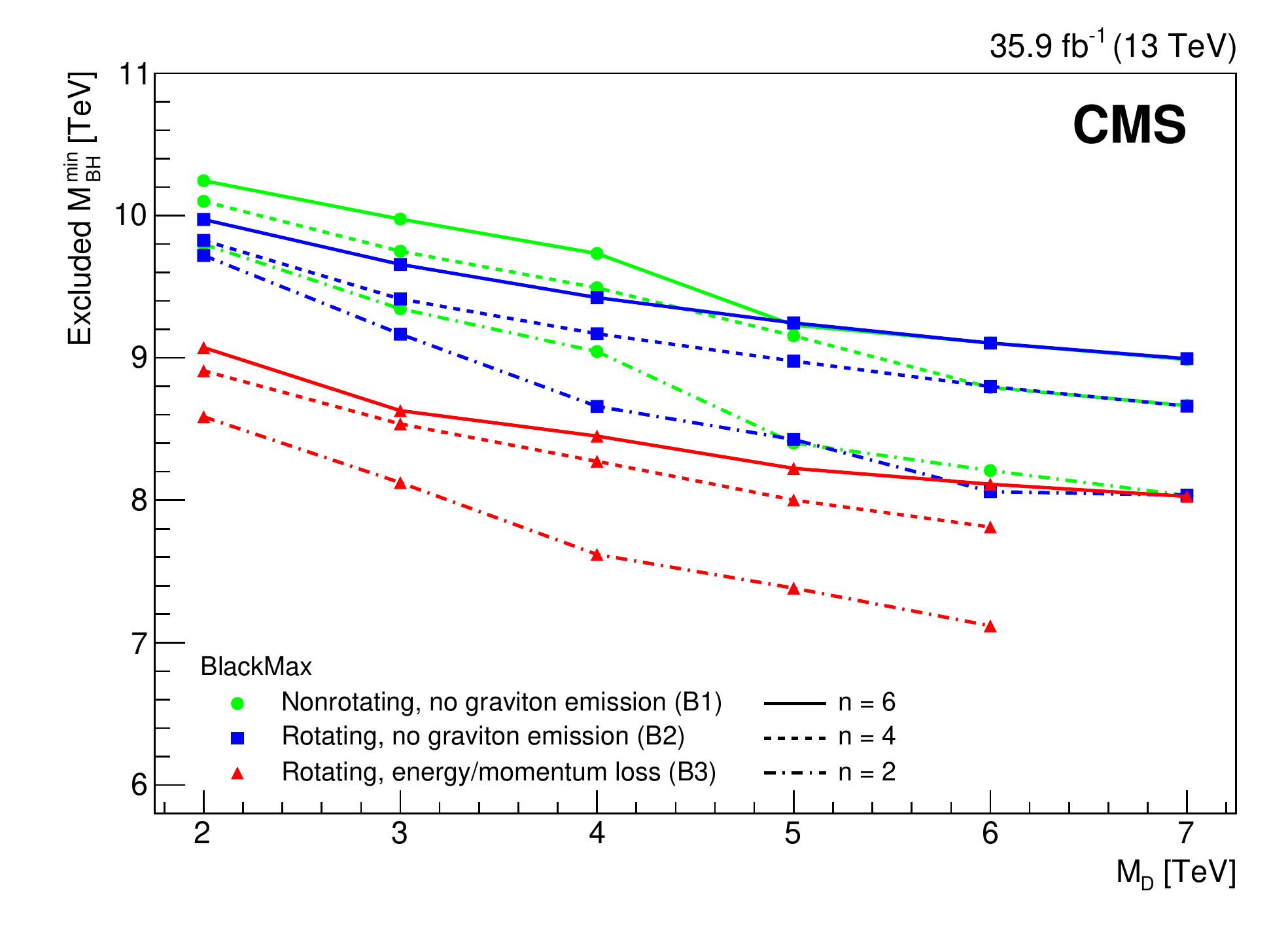}
    \caption{The observed 95\% \CL lower limits on \MBHmin as a function of \MD at different $n$ for the models B1--B3 generated with \BLACKMAX.}
    \label{fig:BlackMax_limit}
\end{figure}

\begin{figure}[htb!]
    \centering
    \includegraphics[height=0.6\textwidth]{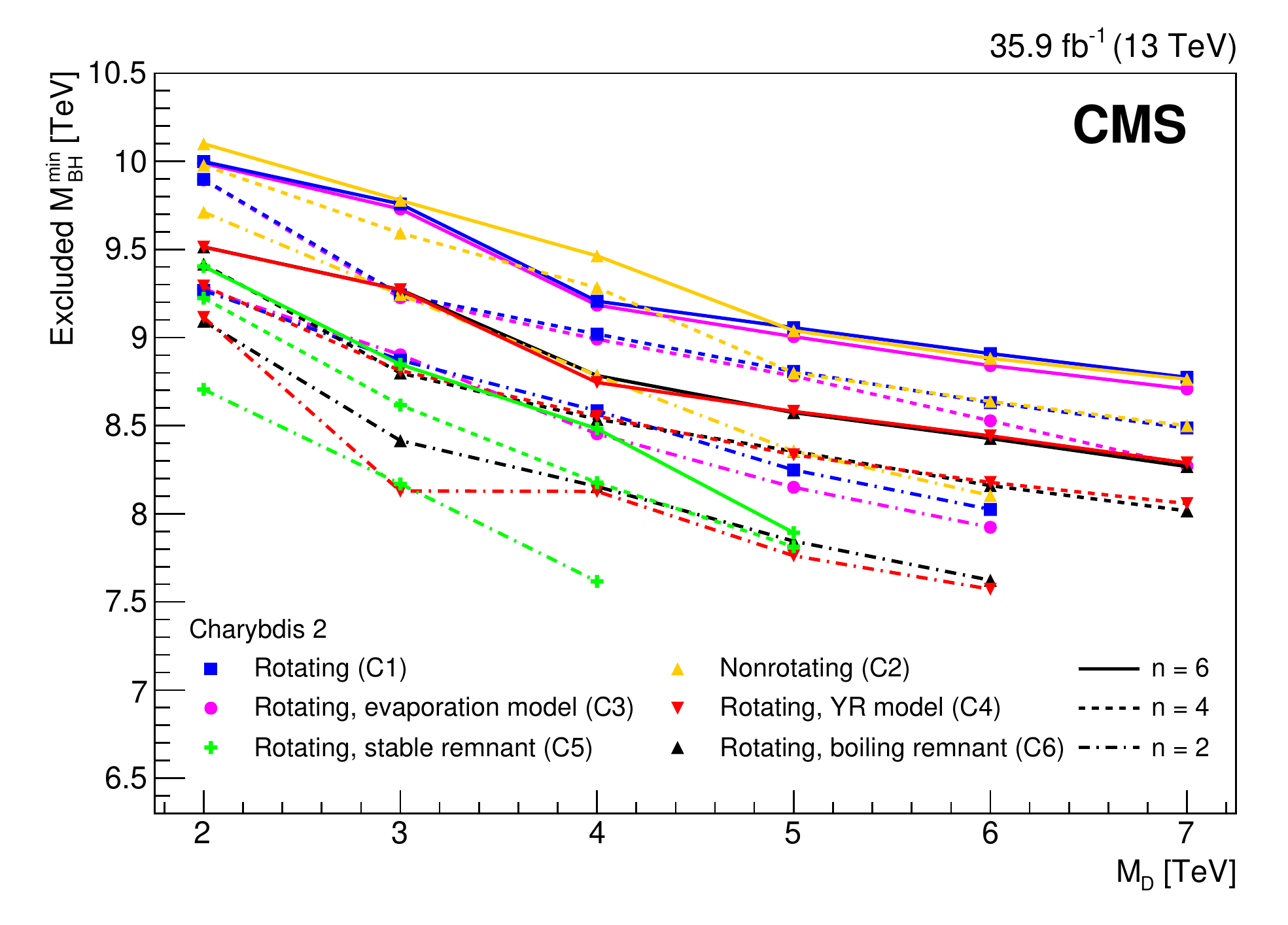}
    \caption{The 95\% observed \CL lower limits on \MBHmin as a function of \MD at different $n$ for the models C1--C6 generated with \CHARYBDIS2.}
    \label{fig:Charybdis_limit}
\end{figure}

\begin{figure}[htb!]
    \centering
    \includegraphics[width=0.49\textwidth]{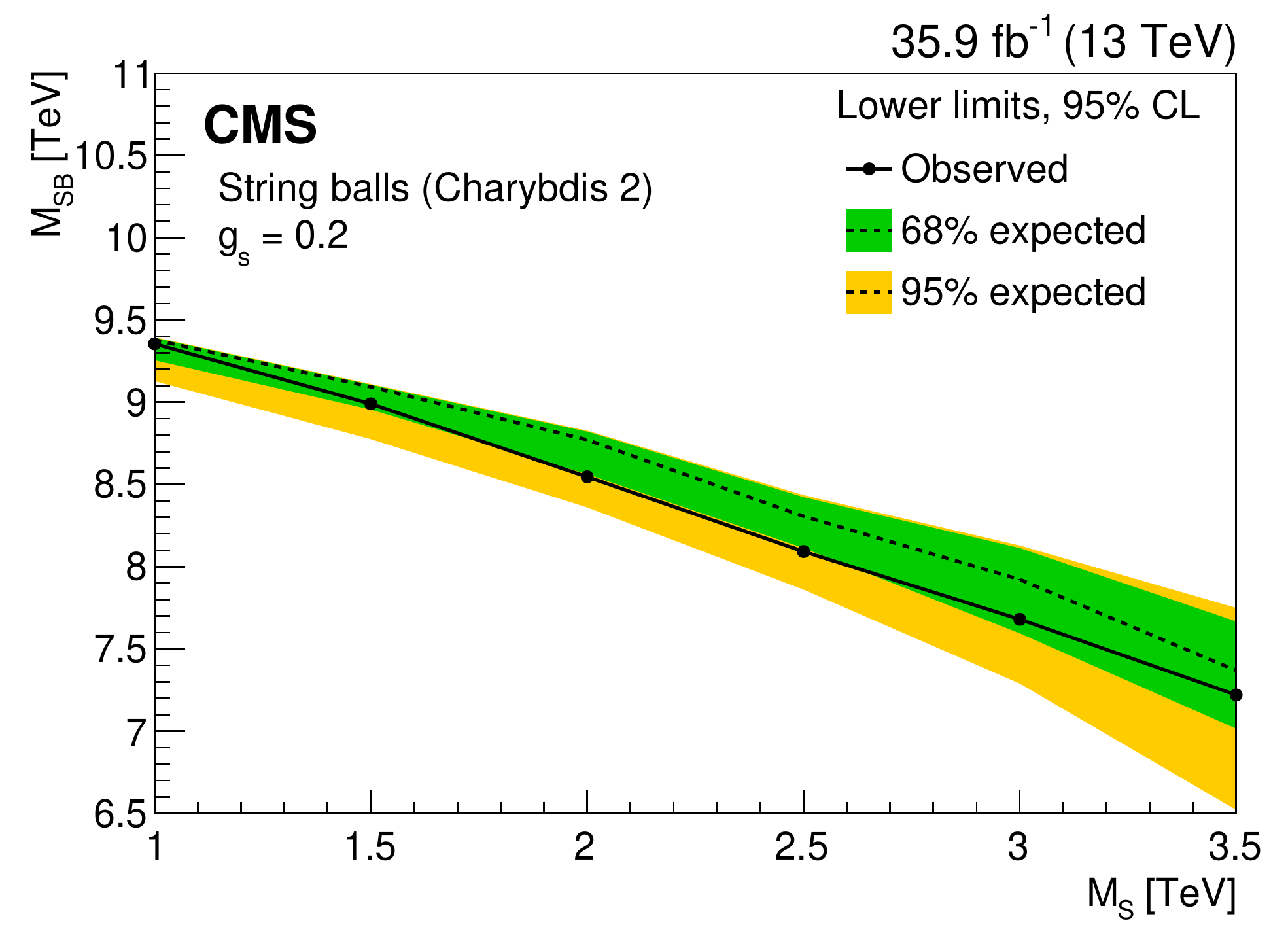}
    \includegraphics[width=0.49\textwidth]{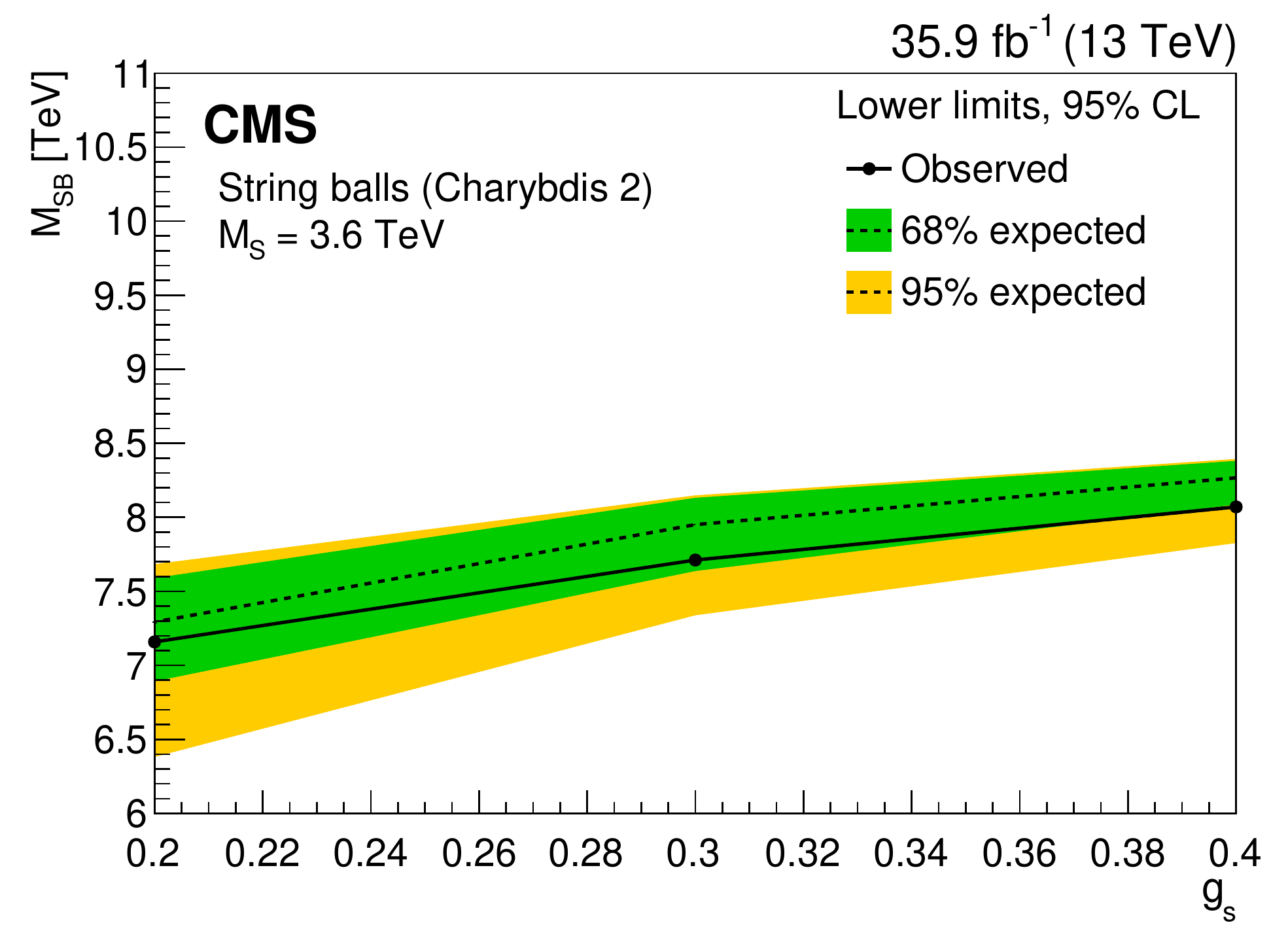}
    \caption{The 95\% \CL lower limits on a string ball mass as a function of the string scale \MS for a fixed value of the string coupling $\gs = 0.2$ (left) and as a function of the string coupling \gs for a fixed value of the string scale $\MS = 3.6\TeV$ (right). The inner (outer) band represents the $\pm 1$ ($\pm 2$) standard deviation uncertainty in the expected limit. The area below the solid curve is excluded by this search.}
    \label{fig:SB}
\end{figure}

We repeat the above procedure for all chosen benchmark scenarios of semiclassical BHs, listed in Tables~\ref{table:generator-blackMax} and \ref{table:generator-Charybdis}. The resulting observed limits on the \MBHmin are shown in Figs.~\ref{fig:BlackMax_limit} and \ref{fig:Charybdis_limit}, for the \BLACKMAX and \CHARYBDIS2 benchmarks, respectively. We also obtain similar limits on the SB mass for the set of the SB model parameters we scanned. These limits are shown in Fig.~\ref{fig:SB} for a fixed string scale $\MS = 3.6\TeV$, as a function of the string coupling \gs (left plot) and for a fixed string coupling $\gs = 0.2$ as a function of the string scale \MS (right plot). The search excludes SB masses below 7.1--9.4\TeV, depending on the values of the string scale and coupling.

\begin{figure}[htb!]
    \centering
    \includegraphics[width=0.6\textwidth]{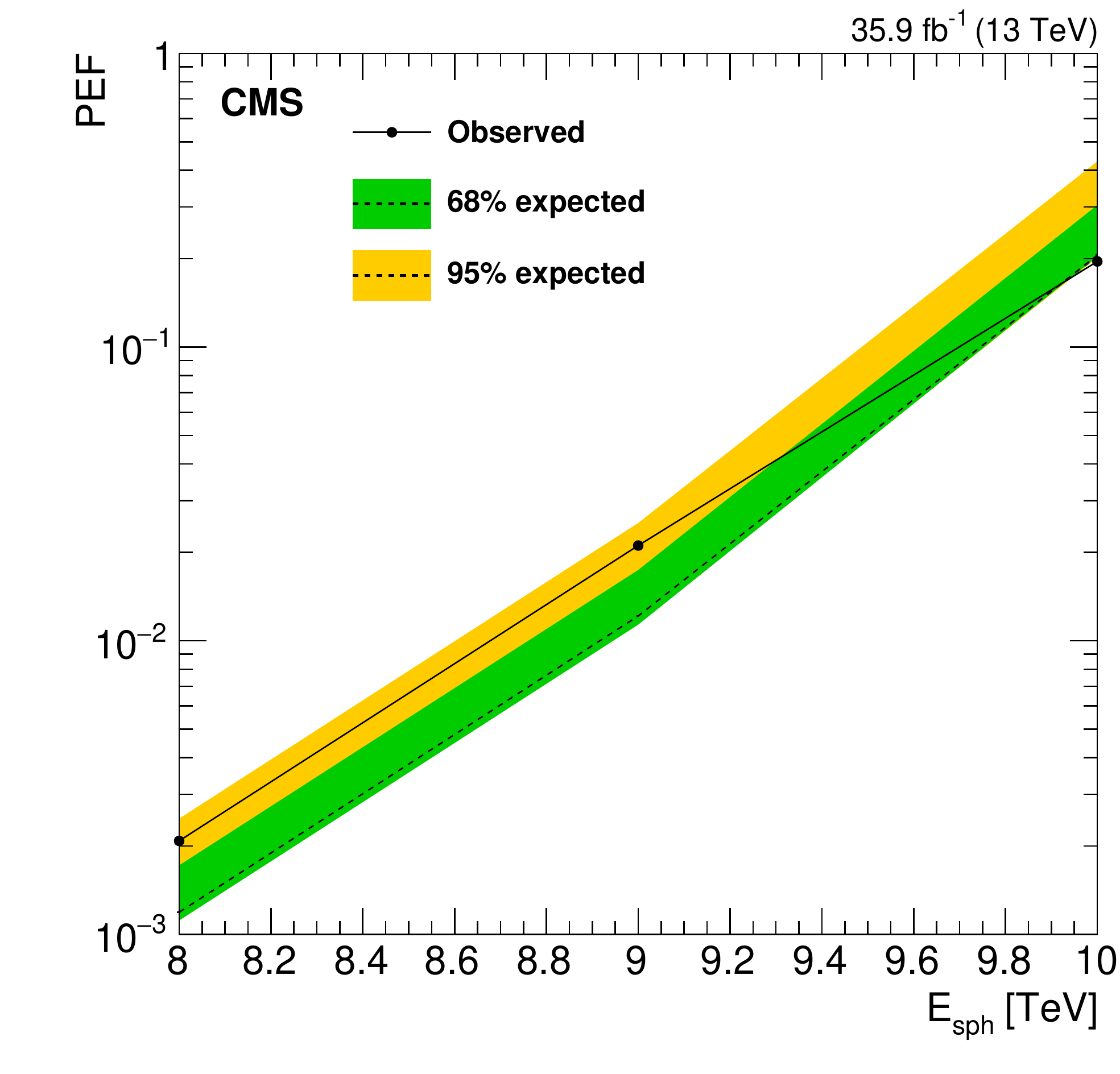}
    \caption{Observed (solid curve) and expected (dashed black curve) 95\% \CL upper limit on the pre-exponential factor PEF of the sphaleron production as a function of \ES. The inner (outer) band represents the $\pm 1$ ($\pm 2$) standard deviation uncertainty in the expected limit. The area above the solid curve is excluded by this search.}
    \label{fig:PEF_limit}
\end{figure}

For the sphaleron signal, the optimal  $(\Nmin,\STmin)$ point is also chosen by scanning for the lowest expected limit and is found
to be $(8,6.2\TeV)$ for $\ES = 9$ and 10\TeV, and $(9,5.6\TeV)$ for $\ES = 8\TeV$. Consequently, the exclusion limit on the sphaleron cross section can be converted into a limit on the PEF, defined in Section~\ref{sec:sphaleron-signal}. Following Ref.~\cite{Ellis2016} we calculate the PEF limits for the nominal $\ES = 9\TeV$, as well as for the modified values of $\ES = 8$ and 10\TeV. The observed and expected 95\% \CL upper limits on the PEF are shown in Fig.~\ref{fig:PEF_limit}. The observed (expected) limit obtained for the nominal $\ES = 9\TeV$ is 0.021 (0.012), which is an order of magnitude more stringent than the limit obtained in Ref.~\cite{Ellis2016} based on the reinterpretation of the ATLAS result~\cite{ATLAS-inclusive13}.

\section{Summary\label{s:summary}}

A search has been presented for generic signals of beyond the standard model physics resulting in energetic multi-object  final states, such as would be produced by semiclassical black holes, string balls, and electroweak sphalerons. The search was based on proton-proton collision data at a center-of-mass energy of 13\TeV, collected with the CMS detector in 2016 and corresponding to an integrated luminosity of \ilum. The background, dominated by QCD multijet production, is determined solely from low-multiplicity samples in data. Comparing the distribution of the total transverse momentum \ST of the final-state objects in data with that expected from the backgrounds, we set 95\% confidence level model-independent upper limits on the product of the production cross section and acceptance for such final states, as a function of the minimum \ST for minimum final-state multiplicities between 3 and 11. These limits reach 0.08\unit{fb} at high \ST thresholds. By calculating the acceptance values for benchmark black hole, string ball, and sphaleron signal models, we convert these model-independent limits into lower limits on the minimum semiclassical black hole mass and string ball mass. The limits extend as high as 10.1\TeV, thus improving significantly on previous results. We have also set the first experimental upper limit on the electroweak sphaleron pre-exponential factor of 0.021 for the sphaleron transition energy of 9\TeV.

\begin{acknowledgments}
We congratulate our colleagues in the CERN accelerator departments for the excellent performance of the LHC and thank the technical and administrative staffs at CERN and at other CMS institutes for their contributions to the success of the CMS effort. In addition, we gratefully acknowledge the computing centers and personnel of the Worldwide LHC Computing Grid for delivering so effectively the computing infrastructure essential to our analyses. Finally, we acknowledge the enduring support for the construction and operation of the LHC and the CMS detector provided by the following funding agencies: BMBWF and FWF (Austria); FNRS and FWO (Belgium); CNPq, CAPES, FAPERJ, FAPERGS, and FAPESP (Brazil); MES (Bulgaria); CERN; CAS, MoST, and NSFC (China); COLCIENCIAS (Colombia); MSES and CSF (Croatia); RPF (Cyprus); SENESCYT (Ecuador); MoER, ERC IUT, and ERDF (Estonia); Academy of Finland, MEC, and HIP (Finland); CEA and CNRS/IN2P3 (France); BMBF, DFG, and HGF (Germany); GSRT (Greece); NKFIA (Hungary); DAE and DST (India); IPM (Iran); SFI (Ireland); INFN (Italy); MSIP and NRF (Republic of Korea); MES (Latvia); LAS (Lithuania); MOE and UM (Malaysia); BUAP, CINVESTAV, CONACYT, LNS, SEP, and UASLP-FAI (Mexico); MOS (Montenegro); MBIE (New Zealand); PAEC (Pakistan); MSHE and NSC (Poland); FCT (Portugal); JINR (Dubna); MON, RosAtom, RAS, RFBR, and NRC KI (Russia); MESTD (Serbia); SEIDI, CPAN, PCTI, and FEDER (Spain); MOSTR (Sri Lanka); Swiss Funding Agencies (Switzerland); MST (Taipei); ThEPCenter, IPST, STAR, and NSTDA (Thailand); TUBITAK and TAEK (Turkey); NASU and SFFR (Ukraine); STFC (United Kingdom); DOE and NSF (USA).

\hyphenation{Rachada-pisek} Individuals have received support from the Marie-Curie program and the European Research Council and Horizon 2020 Grant, contract No. 675440 (European Union); the Leventis Foundation; the A. P. Sloan Foundation; the Alexander von Humboldt Foundation; the Belgian Federal Science Policy Office; the Fonds pour la Formation \`a la Recherche dans l'Industrie et dans l'Agriculture (FRIA-Belgium); the Agentschap voor Innovatie door Wetenschap en Technologie (IWT-Belgium); the F.R.S.-FNRS and FWO (Belgium) under the ``Excellence of Science - EOS" - be.h project n. 30820817; the Ministry of Education, Youth and Sports (MEYS) of the Czech Republic; the Lend\"ulet (``Momentum") Programme and the J\'anos Bolyai Research Scholarship of the Hungarian Academy of Sciences, the New National Excellence Program \'UNKP, the NKFIA research grants 123842, 123959, 124845, 124850 and 125105 (Hungary); the Council of Science and Industrial Research, India; the HOMING PLUS program of the Foundation for Polish Science, cofinanced from European Union, Regional Development Fund, the Mobility Plus programme of the Ministry of Science and Higher Education, the National Science Center (Poland), contracts Harmonia 2014/14/M/ST2/00428, Opus 2014/13/B/ST2/02543, 2014/15/B/ST2/03998, and 2015/19/B/ST2/02861, Sonata-bis 2012/07/E/ST2/01406; the National Priorities Research Program by Qatar National Research Fund; the Programa Estatal de Fomento de la Investigaci{\'o}n Cient{\'i}fica y T{\'e}cnica de Excelencia Mar\'{\i}a de Maeztu, grant MDM-2015-0509 and the Programa Severo Ochoa del Principado de Asturias; the Thalis and Aristeia programmes cofinanced by EU-ESF and the Greek NSRF; the Rachadapisek Sompot Fund for Postdoctoral Fellowship, Chulalongkorn University and the Chulalongkorn Academic into Its 2nd Century Project Advancement Project (Thailand); the Welch Foundation, contract C-1845; and the Weston Havens Foundation (USA).
\end{acknowledgments}

\bibliography{auto_generated}
\cleardoublepage \appendix\section{The CMS Collaboration \label{app:collab}}\begin{sloppypar}\hyphenpenalty=5000\widowpenalty=500\clubpenalty=5000\vskip\cmsinstskip
\textbf{Yerevan Physics Institute, Yerevan, Armenia}\\*[0pt]
A.M.~Sirunyan, A.~Tumasyan
\vskip\cmsinstskip
\textbf{Institut f\"{u}r Hochenergiephysik, Wien, Austria}\\*[0pt]
W.~Adam, F.~Ambrogi, E.~Asilar, T.~Bergauer, J.~Brandstetter, M.~Dragicevic, J.~Er\"{o}, A.~Escalante~Del~Valle, M.~Flechl, R.~Fr\"{u}hwirth\cmsAuthorMark{1}, V.M.~Ghete, J.~Hrubec, M.~Jeitler\cmsAuthorMark{1}, N.~Krammer, I.~Kr\"{a}tschmer, D.~Liko, T.~Madlener, I.~Mikulec, N.~Rad, H.~Rohringer, J.~Schieck\cmsAuthorMark{1}, R.~Sch\"{o}fbeck, M.~Spanring, D.~Spitzbart, A.~Taurok, W.~Waltenberger, J.~Wittmann, C.-E.~Wulz\cmsAuthorMark{1}, M.~Zarucki
% ...
\vskip\cmsinstskip
% ...
1:  Also at Vienna University of Technology, Vienna, Austria\\
% ...
\end{sloppypar}
\end{document}